\documentclass[a4paper,11pt]{article} 
\usepackage{jcappub}
\usepackage{graphicx}
\usepackage{txfonts}
\usepackage{hyperref}
\usepackage{lipsum}
\usepackage{tabularx}
\usepackage{rotating}
\usepackage{comment}
\usepackage{xcolor}
\usepackage{booktabs}
\usepackage{soul}
\usepackage[utf8]{inputenc}
\usepackage{longtable}
\usepackage{caption}
\usepackage{arydshln}
\title{\boldmath Unveiling Electron Density Profile in Nearby Galaxies using SDSS MaNGA}

\author[a]{Shivam Burman,}
\author[b,c]{Sunil Malik\footnote[1]{Corresponding Author},}
\author[a]{Suprit Singh,}
\author[b]{Yogesh Wadadekar}

\affiliation[a]{Indian Institute of Technology Delhi, Hauz Khas, New Delhi, Delhi 110016, India}
\affiliation[b]{National Centre for Radio Astrophysics, Tata Institute of Fundamental Research,  Ganeshkhind, Pune 411007, Maharashtra, India}
\affiliation[c]{Departamento de Física de la Tierra y Astrofísica \& IPARCOS-UCM, Universidad Complutense de Madrid, 28040 Madrid, Spain}
\emailAdd{smalik@ucm.es}

\abstract{Most observational studies of galactic-scale magnetic fields using Faraday rotation rely on estimates of thermal electron densities in galaxies and their radial variations. However, the spatial distribution of electrons in the interstellar medium (ISM) is not clearly known. In this study, we propose and utilize collision-excited doublet emission line ratios of [S\,\textsc{ii}] $\lambda\lambda$ 6716, 6731 \AA~ to estimate the electron densities ($n_e$). To map their distribution in the galaxies, we employ IFU spectroscopic observations from the SDSS MaNGA survey, utilising data products from both the MaNGA Data Analysis Pipeline (DAP). We present a spatially resolved analysis of $66$ face-on galaxies (inclination, $i \leq 15^\circ$), including $46$ star-forming galaxies (SFGs) and $20$ Non-SFGs. Azimuthally averaged radial profiles of $n_e$ are obtained. We found that both SFGs and Non-SFGs exhibit $n_e$ gradients, with higher densities of $n_e$(S\,\textsc{ii}) = $52.87 \pm 8.32$ cm$^{-3}$ and $99.39 \pm 24.37$ cm$^{-3}$, respectively, in the inner disk region (r/R$_e$ $\leq$ 1.5), which decreases to $n_e$(S\,\textsc{ii}) = $20.92 \pm 4.2$ cm$^{-3}$ in SFGs and $34.64 \pm 11.24$ cm$^{-3}$ in Non-SFGs, in the outer disk region (r/R$_e$ $>$ 1.5). We have also analysed these sources with \texttt{Pipe3D} fluxes. We translated $n_e$ to electron column densities ($N_e$) by assuming a typical disk of thickness 1 kpc and note that $N_e \sim 10^{22}$ cm$^{-2}$ at $\sim$14 kpc in the disk outer region. We have also discussed the profiles obtained using [O\,\textsc{ii}] $\lambda\lambda$ 3726, 3729 \AA \ doublet. These electron density estimates at different radii provide valuable insights for resolving ambiguities in current and future studies of magnetic fields in galaxies. 
}

\begin{document}
\maketitle{}
\flushbottom

\section{Introduction} \label{sec:intro}

Electron density is a key parameter that specifies the physical state of the interstellar medium (ISM) and the circumgalactic medium (CGM) of the galaxies. Several physical observables, for example, the metal abundances in galaxies \cite{Osterbrock_book_2006}, galactic magnetism, and pulse dispersion \cite{Snell_2019} depend on the electron density. Specifically, the thermal electron column density ($N_e$) which is the line integrated $n_e$ goes into determining the galactic magnetic fields strengths (see \cite{Bernet2008,bernet_extent_2013,Malik_2020}) as well as its radial profile \cite{Burman_2024} using Faraday Rotation. The studies have largely assumed a homogeneous $N_e$ due to a lack of any significant observations of the spatial distribution of electron densities in the ISM and the CGM. This largely unexplored gap motivates us to investigate the variation of $n_e$ in the spatially as well as spectroscopically well-resolved galaxies in the SDSS Mapping Nearby Galaxies at Apache Point Observatory (MaNGA) data.

For our own galaxy, the NE2001 \cite{NE2001} and YMW16 \cite{Yao_2017} models have used dispersion measures (DM) from pulsars in the Milky Way and the Magellanic Clouds, as well as extragalactic sources, including Fast Radio Bursts (FRBs), to estimate the electron densities and build its profile. Similarly, the DMs of the FRBs offer a means to probe the electron densities of their host galaxies and any intervening galaxies (\cite{Petroff2019}). The observed DM is influenced by the contributions from the Milky Way and the intergalactic medium (IGM), necessitating accurate models of these components to segregate the host contribution. Nevertheless, these observations provide electron density estimates only at specific radii given by the line of sight, and cannot give the complete picture of the spatial variation of the $n_e$ in the host. We, therefore, look into other probes.

Emission lines such as [O\,\textsc{ii}] $\lambda\lambda$3726, 3729 \AA, [S\,\textsc{ii}] $\lambda\lambda$6716, 6731 \AA, [N\,\textsc{ii}] 122 and 205 $\mu$m fine-structure line emission have been used widely to probe the electron densities in galaxies. These lines are associated with active star formation, and the flux ratio of the doublets is directly connected to the electron density through the physics of collisional excitation and de-excitation. The average electron density in star-forming galaxies (SFGs) has been observed to evolve with redshift, ranging from 200-300 $\text{cm}^{-3}$ at $z\sim$ 2-3 \cite{Steidel_2014,Shimakawa_2015, Sanders_2016, Davies_2021}, to 100-200 $\text{cm}^{-3}$ at $z\sim$ 1.5 \cite{Kaasinen_2017,Kashino_2017,Harshan_2020}, and further decreasing to $\sim$32 $\text{cm}^{-3}$ in the local galaxies \cite{Herrera-Camus_2016,Sanders_2016,Kashino_2019}. Most of these studies give estimates of average $n_e$ using single-fiber spectroscopy which captures the integrated light from the entire galaxy without resolving its spatial structure. Although these give valuable insights about the overall ionised gas properties, our interest lies in resolving the spatial distribution of thermal electrons. As remarked earlier, this is particularly relevant for magnetic field studies utilising Faraday Rotation.

The integral field spectroscopic (IFS) surveys provide a crucial tool for studying spatially resolved gas properties over larger distances (\cite{Law_2021, Belfiore_2022}). Recently, \cite{Espinosa_CALIFA_2022} (hereafter EP22) analysed the spatially resolved properties of ~924 galaxies from the Calar Alto Legacy Integral Field Area (eCALIFA) survey (see \cite{CALIFA_survey_2012,Espinosa_Ponce_2020}). Additionally, \cite{Barrera2023sdss} (hereafter BB23) utilised data cubes from the SDSS MaNGA survey \cite{Bundy_2015,Yan_2016}. Their sample selection criteria, that the galaxies have a reliable coverage, a consistent physical spatial resolution, and that the evolution of the galaxies is similar, selected roughly 13\% of the total MaNGA dataset. The resulting profiles suggest mildly decreasing radial trends of the electron density in certain cases. In this paper, we probe the radial profiles of $n_e$ by utilising much more stringent criteria of face-on galaxies as well as a classification of the galaxies based on their star-forming activity, visual checks, and removal of AGNs, to have a homogeneous selection. The electron densities are estimated using the line ratios of the [S\,\textsc{ii}] $\lambda\lambda$6716, 6731 \AA~doublet and the face-on criterion allows an unambiguous translation of the electron density to the electron column density. For completeness, we also provide the radial profiles derived using [O\,\textsc{ii}] $\lambda\lambda$ 3726, 3729 \AA \ doublet in Appendix~\ref{appendix:appendix_oii}. 

This paper is structured as follows: In Section \ref{sec:data_and_sample}, by briefly introducing the SDSS MaNGA survey, we give the details of our sample creation. Section \ref{sec:analysis_and_results} discusses the methodology employed in estimating the electron densities for our sample dataset. We build the individual as well as co-added radial profiles of electron density for various cases. A discussion of our results is given in Section \ref{sec:discussion} in comparison with the existing literature. Finally, we present the conclusions of our study in Section \ref{sec:conclusion}.

In this paper, we have used \texttt{LambdaCDM} cosmological model with the cosmological parameters utilised in this model are $H_0 = 67.74$ km s$^{-1}$ Mpc$^{-1}$, $\Omega_m = 0.31$, and $\Omega_\Lambda = 0.69$ \citep{Planck_2015}.

\section{DATA AND SAMPLE} \label{sec:data_and_sample}

To investigate the extent and structure of the electron density in the galaxies, we have used data releases from the IFS surveys. In several past studies \cite{Cappellari_Atlas_2011, Croom_2012, CALIFA_survey_2012, Brodie_2014, Ma_2014}, authors have used different IFU surveys ATLAS$^{3\text{D}}$, SAMI, SLUGGS, MASSIVE, and CALIFA \footnote{\href{https://www-astro.physics.ox.ac.uk/atlas3d/}{ATLAS$^{3\text{D}}$}, \href{https://sami-survey.org/}{SAMI}, \href{https://sluggs.swin.edu.au/Start.html}{SLUGGS}, \href{https://blackhole.berkeley.edu/}{MASSIVE}, \href{https://califa.caha.es/}{CALIFA}}, having different characteristics and source properties. In this study, we have used the data from the SDSS MaNGA survey, which is the largest existing IFS survey with $\sim$10000 nearby galaxies. The MaNGA survey has been chosen for two main reasons: (i) it provides a continuous coverage of wavelength that allows us full range model fitting for many spectral features, (ii) the spectral resolution of MaNGA is $R\sim2000$ at the average redshift $\left(\langle z \rangle \sim 0.03\right)$. 

\newpage
\setlength{\LTcapwidth}{\textwidth}
\captionsetup{width=\textwidth}      

\begin{longtable}{|@{\extracolsep{\fill}}c|c|c|c|c|c|c|}
\hline \hline
MaNGA ID & RA & DEC & $z$ & $\text{R}_e$ & $i$ & $\text{sSFR}$ \\
 & (degrees) & (degrees) & & (arcsec) & (degrees) & ($\log \text{yr}^{-1}$) \\ \hline \hline

    1-38014 & 49.821442 & -0.969631 & 0.0538 & 3.1067 & 14.7907 & -9.7324 \\
    1-41843 & 29.984667 & 12.459571 & 0.0388 & 3.523 & 13.2803 & -9.7852 \\ 
    1-404900 & 185.326001 & 32.711062 & 0.023 & 5.7845 & 13.548 & -9.8021 \\
    1-22383 & 253.31305 & 64.476279 & 0.0542 & 2.9758 & 14.0652 & -9.9238 \\
    1-51001 & 48.154706 & -7.572848 & 0.036 & 3.6243 & 12.4599 & -9.9241 \\ 
    1-603309 & 32.88984 & 13.917127 & 0.0265 & 3.0718 & 9.2752 & -9.9724 \\
    1-247976 & 239.436638 & 41.709524 & 0.0238 & 3.3666 & 10.7232 & -10.0087 \\
    1-35275 & 30.276873 & -0.651504 & 0.0429 & 3.7128 & 7.9286 & -10.0291 \\
    1-462994 & 135.365943 & 21.175119 & 0.0255 & 7.4548 & 13.3968 & -10.0509 \\
    1-39661 & 14.779524 & 14.523282 & 0.0474 & 6.7369 & 13.3434 & -10.0532 \\
    1-97120 & 316.262701 & -6.57704 & 0.0406 & 2.3092 & 10.6469 & -10.1046 \\
    1-40792 & 17.775768 & 15.172137 & 0.0382 & 8.1523 & 4.1232 & -10.116 \\
    1-609044 & 156.727477 & -0.541518 & 0.0346 & 3.3917 & 9.5636 & -10.1186 \\
    1-45151 & 122.79107 & 45.66357 & 0.0229 & 9.3223 & 13.2924 & -10.142 \\
    1-246191 & 220.043768 & 53.1002 & 0.0367 & 2.4842 & 13.3786 & -10.1451 \\
    1-38894 & 54.736378 & 0.917886 & 0.0417 & 5.6715 & 12.7331 & -10.1519 \\
    1-595900 & 311.75303 & -6.16493 & 0.0224 & 5.8987 & 14.9974 & -10.1596 \\
    1-113404 & 316.882257 & 10.977209 & 0.0423 & 3.5517 & 12.6928 & -10.167 \\
    1-41133 & 20.528024 & 15.570976 & 0.0375 & 2.5725 & 11.2527 & -10.1688 \\
    1-606221 & 148.54456 & 2.28718 & 0.0246 & 14.3579 & 9.1565 & -10.2005 \\
    1-624369 & 196.80506 & 28.04697 & 0.0244 & 7.1805 & 13.0121 & -10.2169 \\
    1-115262 & 334.060494 & 12.395525 & 0.0679 & 1.824 & 12.266 & -10.2354 \\
    1-248037 & 239.102072 & 42.395486 & 0.0407 & 5.8938 & 11.9708 & -10.2654 \\
    1-324250 & 249.534262 & 28.368678 & 0.0738 & 5.1779 & 10.1798 & -10.2667 \\
    1-593748 & 229.32448 & 29.40054 & 0.0175 & 15.0147 & 14.5972 & -10.2902 \\
    1-43865 & 118.317318 & 39.050123 & 0.0409 & 5.9905 & 11.9085 & -10.291 \\
    1-61876 & 163.588577 & 3.395789 & 0.0349 & 2.4455 & 13.3745 & -10.3053 \\
    1-322277 & 229.164859 & 42.953852 & 0.018 & 12.8781 & 13.4192 & -10.3106 \\
    1-63637 & 178.713766 & 2.959228 & 0.0202 & 10.3863 & 14.9645 & -10.3222 \\
    1-32827 & 14.560282 & -0.296829 & 0.0444 & 5.4253 & 13.3697 & -10.3249 \\
    1-234408 & 199.94477 & 48.206576 & 0.036 & 6.6171 & 14.7096 & -10.3541 \\
    1-495795 & 178.76449 & 24.012522 & 0.0771 & 4.128 & 12.8222 & -10.3645 \\
    1-623940 & 195.657162 & 50.438429 & 0.0236 & 14.9886 & 11.2829 & -10.3907 \\
    1-117697 & 348.146861 & 13.27445 & 0.067 & 5.5397 & 5.4809 & -10.4024 \\
    1-196310 & 204.968747 & 52.386534 & 0.0583 & 5.1591 & 12.6103 & -10.4025 \\
    1-594816 & 254.623324 & 20.041471 & 0.0201 & 15.5174 & 10.3451 & -10.4239 \\
    1-230291 & 126.384623 & 27.608709 & 0.0196 & 3.9665 & 12.5932 & -10.5408 \\
    1-248415 & 241.84668 & 41.70897 & 0.0181 & 4.3563 & 11.1159 & -10.5424 \\
    1-22301 & 253.405559 & 63.03127 & 0.1052 & 4.6054 & 14.1349 & -10.5441 \\
    1-121075 & 114.695391 & 29.891283 & 0.0979 & 2.6594 & 14.7459 & -10.6025 \\
    1-352635 & 126.132758 & 54.85389 & 0.0249 & 14.888 & 13.1024 & -10.6183 \\
    1-385781 & 131.937672 & 25.930421 & 0.0653 & 5.9564 & 10.7473 & -10.6218 \\
    1-55194 & 146.151861 & 2.400868 & 0.0864 & 3.9565 & 7.391 & -10.6615 \\
    1-301537 & 145.860507 & 36.237381 & 0.0218 & 5.8276 & 9.8223 & -10.6932 \\
    1-182358 & 150.739859 & 6.271636 & 0.0215 & 9.6537 & 7.3078 & -10.7478 \\ 
    1-90849 & 237.582743 & 56.131981 & 0.0661 & 5.0286 & 13.4553 & -10.7634 \\ \hdashline
    1-575738 & 204.111415 & 43.01508 & 0.062 & 8.1864 & 12.8043 & -11.8082 \\
    1-31500 & 7.122727 & 0.78103 & 0.0403 & 4.9062 & 12.3448 & -11.859 \\
    1-263694 & 233.654226 & 30.582284 & 0.0555 & 6.6572 & 12.2432 & -11.9108 \\
    1-576031 & 180.980701 & 50.872723 & 0.0714 & 7.3986 & 14.3477 & -11.9373 \\
    1-190110 & 187.841004 & 53.875684 & 0.1452 & 3.5919 & 14.2899 & -11.9667 \\
    1-586585 & 154.858449 & 44.404681 & 0.0234 & 9.331 & 14.5636 & -12.143 \\
    1-28665 & 346.647274 & -0.421577 & 0.0269 & 4.5148 & 14.816 & -12.2019 \\
    1-44726 & 120.715023 & 46.157033 & 0.0611 & 2.9553 & 13.6834 & -12.2331 \\
    1-176422 & 254.054374 & 33.590584 & 0.056 & 2.7764 & 11.3111 & -12.24 \\
    1-337227 & 239.477364 & 9.40992 & 0.0428 & 4.576 & 12.0521 & -12.2521 \\
    1-181975 & 147.512787 & 6.175205 & 0.0616 & 2.3037 & 14.7941 & -12.2775 \\
    1-415592 & 198.470946 & 34.124767 & 0.0373 & 4.3262 & 12.846 & -12.2786 \\
    1-195465 & 195.119569 & 55.194222 & 0.0642 & 2.2813 & 12.7859 & -12.2979 \\
    1-210385 & 245.338708 & 40.309983 & 0.0333 & 6.504 & 14.2651 & -12.3239 \\
    1-364507 & 227.739921 & 7.605085 & 0.0484 & 3.3319 & 14.9211 & -12.3596 \\
    1-60349 & 149.139464 & 1.849215 & 0.0328 & 6.1621 & 10.9177 & -12.4004 \\
    1-570537 & 352.625856 & 13.905628 & 0.0405 & 5.7626 & 13.7577 & -12.4876 \\
    1-76488 & 131.127935 & 2.294926 & 0.0506 & 3.9772 & 12.8088 & -12.6603 \\
    1-631910 & 227.390701 & 7.515313 & 0.0778 & 2.8968 & 12.6459 & -12.7837 \\ \hline

\hline
\caption{The table shows the properties of all 66 galaxies from our sample. The horizontal dashed black line shows the demarcation between the SFGs (above) and Non-SFGs (below).}
\label{tab:table_1}
\end{longtable}

\subsection{The MaNGA survey}
\label{sub:manga_survey}

The MaNGA survey is one of the three major programs of the fourth-generation SDSS-IV \cite{Bundy_2015}. It uses the 2.5m Sloan Foundation Telescope \cite{Gunn_2006} at Apache Point Observatory (APO) to observe a sample $\sim 10000$  galaxies comprised of all morphological types. The targets cover an almost flat distribution in stellar mass (M$_*$) with a range of $10^9 \text{M}_\odot<\text{M}_*<10^{11}\text{M}_\odot$, and a redshift interval of $0.01 \leq z \leq 0.15$ \cite{Blanton_2017}. MaNGA uses the IFU spectroscopy technique, observing each galaxy with 19-127 hexagonal fiber bundles, corresponding to a diameter between 12-32 arcsec in the sky. The two dual-channel Baryon Oscillation Spectroscopic Survey (BOSS) spectrographs \cite{Smee_2013} provide simultaneous spectra that have a wavelength coverage range of 3600-10300 \AA~ with a median spectral resolution of $R \sim 2000$. More details on the observing strategy, calibration, and survey design of MaNGA can be found in \cite{Law_2015}, \cite{Yan_2016}, and \cite{Wake_2017}.

Two-thirds ($\sim$5000) of the MaNGA galaxies have a field of view covering 1.5 times the effective radius (R$_e$; which contains half of the galaxy light in $r$-band) and one-third ($\sim$3300) covers out to 2.5R$_e$ and more. These are referred to as the primary and secondary samples, respectively. A Colour-Enhanced sample of an additional $\sim$1700 galaxies are added to the primary sample to balance the color
distribution at fixed $\text{M}_*$. Hence, this collective sample is known as ``primary+" sample. The primary+ sample has an average redshift of $\langle z \rangle \sim 0.03$, while the secondary sample has a higher average redshift of $\langle z \rangle \sim 0.045$.

We have utilised the data products of the MaNGA Data Analysis Pipeline (DAP; \cite{Westfall_2019,Belfiore_2019}) and \texttt{pyPipe3D} pipeline (hereafter \texttt{Pipe3D}) \cite{LACERDA2022101895,Sánchez_2022}. In both approaches, the stellar continuum is subtracted from the observed spectra and the pure gas spectra are obtained. The algorithm searches for emission line features in the resultant spectra and yields the various physical properties derived from spectroscopic analysis.

\subsection{Sample selection}
\label{sub:sample}
In order to obtain the radial profile of electron density, we require the emission-line fluxes of certain spectral lines. Since the observed line fluxes represent the integrated emission along the line of sight, a face-on orientation minimizes projection effects and internal dust extinction, thereby allowing a better interpretation of the radial distribution \cite{Zhang_2016,Sami_2018}). Thus, we focus only on the face-on galaxies for our analysis. The properties of the galaxies are obtained from the catalog provided by \cite{Sánchez_2022}. The authors obtained the galaxies' properties such as Right Ascension (RA), Declination (DEC), $z$, R$_e$ and inclination ($i$) from the NASA Sloan Atlas (NSA)\footnote{\href{http://www.nsatlas.org/}{NASA-Sloan Atlas}}. We begin by selecting the galaxies with low-inclination angles, $i\leq 15^{\circ}$; this condition yields a sample of 191 face-on galaxies. We removed one galaxy from our sample due to the contamination from a foreground star. In order to categorise the remaining sample of 190 galaxies into star-forming galaxies and other types, we adopt the criteria discussed in \cite{Biswas_2024} and \cite{Salim_2014} based on specific star formation rate (sSFR; defined as the ratio of star formation rate (SFR) and stellar mass), according to which the star-forming galaxies follow,
\begin{equation}
    \text{sSFR} ~[\text{log}(\text{yr}^{-1})] ~\geq ~-10.8
\end{equation} 
whereas the galaxies with
\begin{equation}
    \text{sSFR} ~[\text{log}(\text{yr}^{-1})] ~\leq ~-11.8
\end{equation}
are quenched galaxies and the galaxies that fall between these two are called green-valley galaxies (\cite{Salim_2014}). The star formation rate is estimated by using the dust-corrected $\text{H}\alpha$ luminosities (see \cite{Xu_SFG_S0_2022} and \cite{Sánchez_2022}).

This criterion yields 59 SFGs, 35 green valley galaxies and 96 quenched galaxies (Non-SFGs, hereafter). In this work, we have only utilised the SFGs and Non-SFGs for further analysis. We also removed two active galactic nuclei (AGN) candidates to avoid any contamination from higher densities of the nuclear region of the AGN. Additionally, to obtain a larger radial coverage, we have selected the galaxies with $\text{R}_e > 1.5$ kpc. This reduces the sample to 120 galaxies with 47 SFGs and 73 Non-SFGs. The sample has been further filtered after the flux analysis of the S\,\textsc{ii} and O\,\textsc{ii} doublets, as discussed in detail in subsection~\ref{subsubsec:flux_analysis}. Briefly, we removed the pixels with negative flux values and radially binned the flux maps. We removed those bins which have a negative pixel coverage of more than 30\%. These criteria removed one SFG and 53 Non-SFGs from the sample. This yields the final sample with 46 SFGs and 20 Non-SFGs. The details of all the 66 galaxies are provided in Table~\ref{tab:table_1}.

\begin{figure*}
    \centering
    \includegraphics[
        width=\textwidth,
        trim=1cm 0cm 2cm 0cm,
        clip
    ]{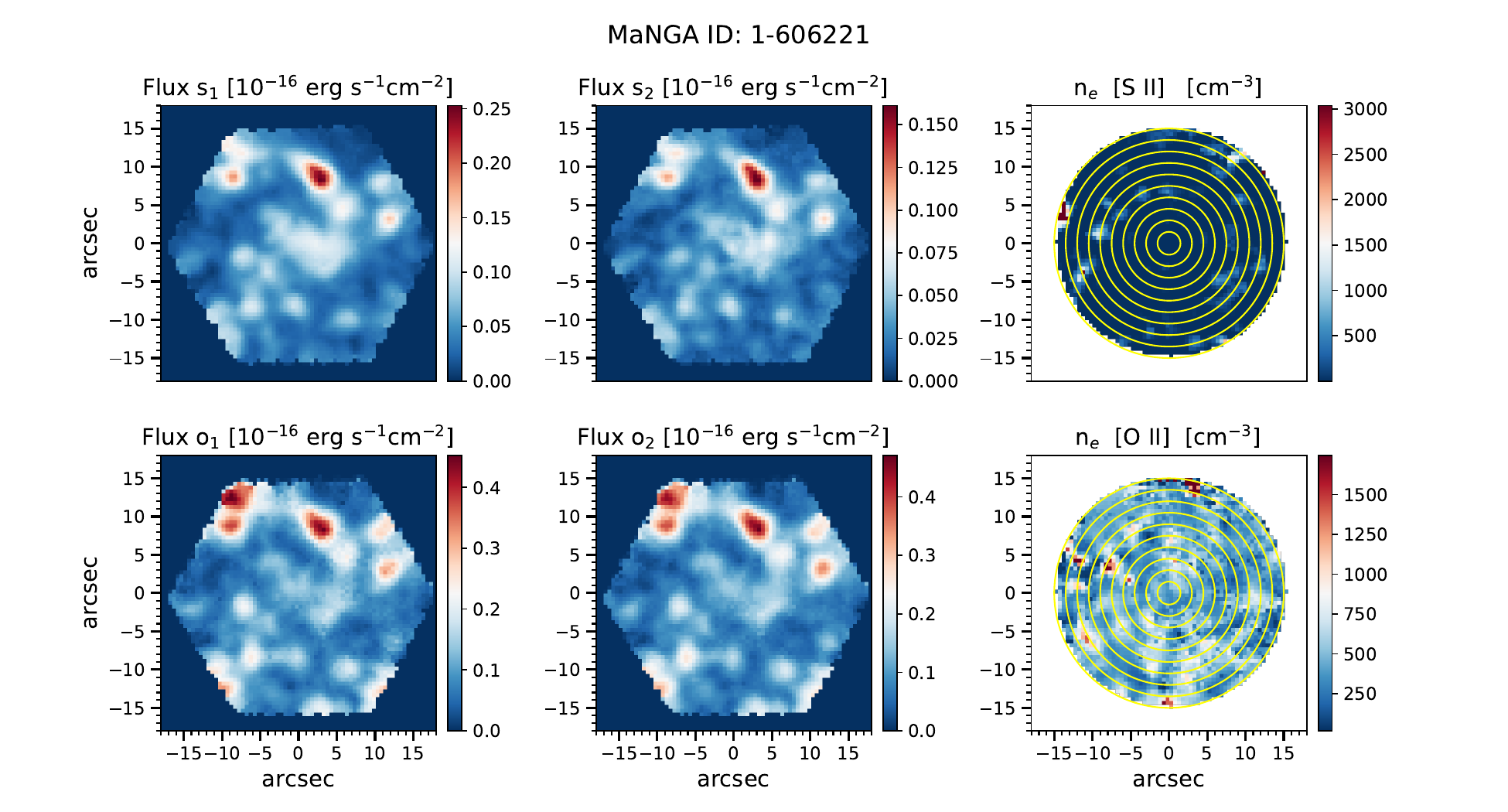}
    \caption{\textit{Upper Panel:} The figure presents the spatial distribution of DAP fluxes s$_1$ (left) and s$_2$(middle), along with the electron density $n_e$ (right), derived from s$_1$ and s$_2$, for the galaxy with "MaNGA ID 1-606221". \textit{Lower Panel:} Similarly, the left and middle panel displays the spatial maps of fluxes o$_1$ and o$_2$, respectively, while the right panel shows the corresponding $n_e$, obtained using o$_1$ and o$_2$. The yellow annuli on both electron density maps represent the linear binning scheme.}
    \label{fig:single_galaxy}
\end{figure*}

\section{ANALYSIS AND RESULTS}
\label{sec:analysis_and_results}

\subsection{Electron Density Diagnostics}
\label{subsec:elec_den_theory}
The electron density in a gas cloud or an H\,\textsc{ii} region can be determined by analysing the effects of collisional excitation and de-excitation. This is achieved by comparing the intensities of two lines from the same ion originating from closely lying energy levels. Since the excitation rates to these levels are primarily governed by the ratio of their collision strengths, differences in their radiative transition probabilities or collisional de-excitation rates cause the relative populations of the levels to vary with density. Consequently, the ratio of the emitted line intensities becomes a reliable diagnostic for determining the electron density \cite{Osterbrock_book_2006,dopita2013astrophysics}. There are several doublets that fall in this regime and lie in a wide optical range, such as [O\,\textsc{ii}] $\lambda\lambda$3726, 3729 \AA, [S\,\textsc{ii}] $\lambda\lambda$6716, 6731 \AA, [Ar\,\textsc{iv}] $\lambda\lambda$4711, 4740 \AA~and [Cl\,\textsc{iii}] $\lambda\lambda$5517, 5537 \AA~(\cite{Kewley_2019}). In this work, we primarily focus on electron densities derived from Gaussian-fitted emission-line fluxes. Since due to the unavailability of the [Ar\,\textsc{iv}] $\lambda\lambda$4711, 4740 \AA~and [Cl\,\textsc{iii}] $\lambda\lambda$5517, 5537 \AA~ fluxes in both DAP and \texttt{Pipe3D} pipelines, we have not included these doublets in our analysis.
Therefore, we use the commonly used diagnostic doublets: [S\,\textsc{ii}] $\lambda\lambda$6716, 6731 \AA, hereafter referred to as s$_1$ (6716) and s$_2$ (6731), (and [O\,\textsc{ii}] $\lambda\lambda$3726, 3729 \AA~, hereafter referred to as o$_1$ (3726) and o$_2$ (3729), in Appendix~\ref{appendix:appendix_oii}), to estimate the electron densities. We adopt the method from \cite{Sanders_2016} (hereafter S15) to derive the electron density using the line ratios described as,

\begin{equation}
    n_e(\text{R}) = \frac{c{\text{R}} - ab}{a - {\text{R}}}, \quad \delta n_e  = \frac{|ac - ab|}{(a - {\text{R}})^2} \delta {\text{R}}
    \label{eqn:equation_ne}
\end{equation}
where $n_e$ is the electron density in $\text{cm}^{-3}$, R = s$_1$/s$_2$ (hereafter R$_{\text{S\,\textsc{ii}}}$) or o$_2$/o$_1$ (hereafter R$_{\text{O\,\textsc{ii}}}$) is the line ratio of the fluxes, $\delta n_e$ is the error in $n_e$ obtained using the propagation of errors, while $a$, $b$ and $c$ are the coefficients that best fit the solutions for the relative populations of the doublets. We adopted the values of a, b and c as 0.4315, 2107 and 627.1 for S\,\textsc{ii}, and 0.3771, 2468 and 638.4 for O\,\textsc{ii} following \cite{Sanders_2016}. These values are based on the atomic data for transition probabilities from \cite{fischer_tachiev_mchf} and the collision strengths for S\,\textsc{ii} and O\,\textsc{ii} are obtained from \cite{Tayal_2010}. Since the doublet lines are very close in wavelength, the fluxes need no correction for dust extinction (\cite{Kaasinen_2017}). 

\begin{figure*}
    \centering
    \includegraphics[scale=0.5]{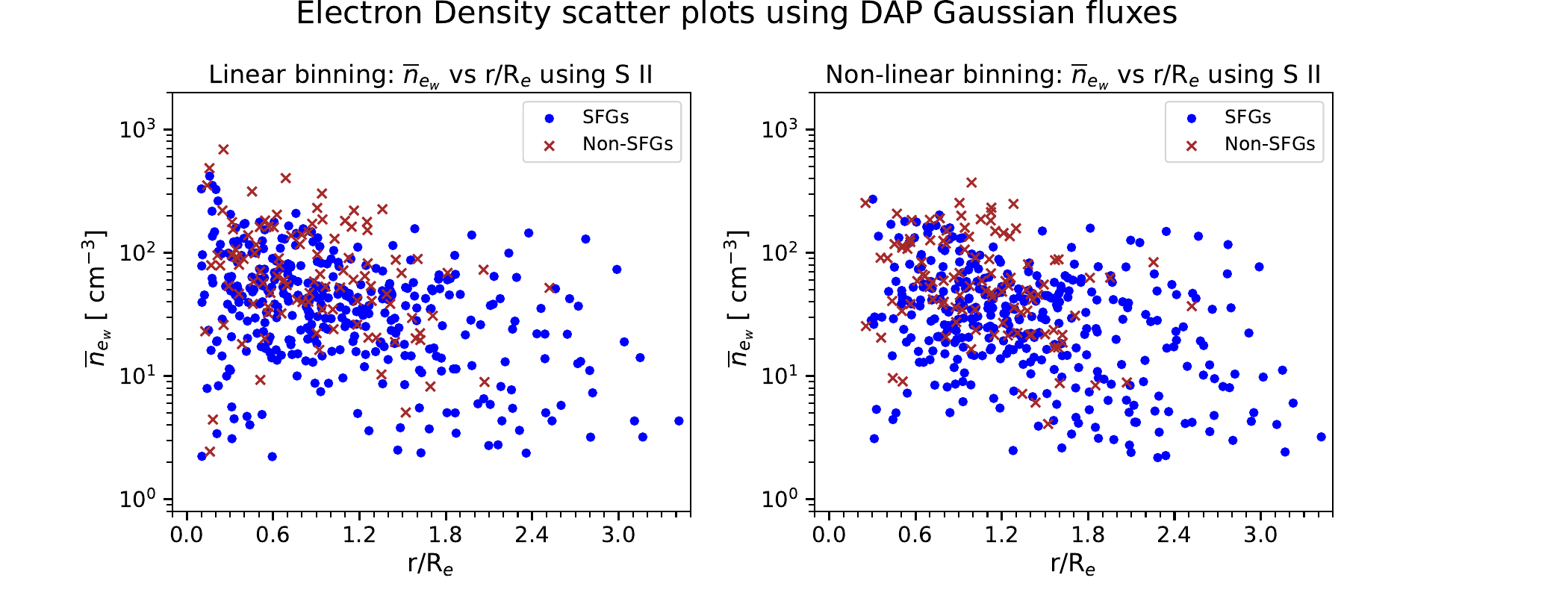}
    \caption{\textit{Left Panel:} The figure illustrates the results from linear binning applied to individual galaxies, using DAP data. The left panel depicts scatter plots of $n_e$(S\,\textsc{ii}) for SFGs (blue dots) and Non-SFGs (brown crosses). Each data point corresponds to the $\bar{n}_{e_w}$ computed across different bins within individual galaxies. \textit{Right Panel:} Similarly, the right panel represent the scatter plots for the case of non-linear binning.}
    \label{fig:figure_2}
\end{figure*}

\begin{figure*}
    \centering
    \includegraphics[scale=0.5]{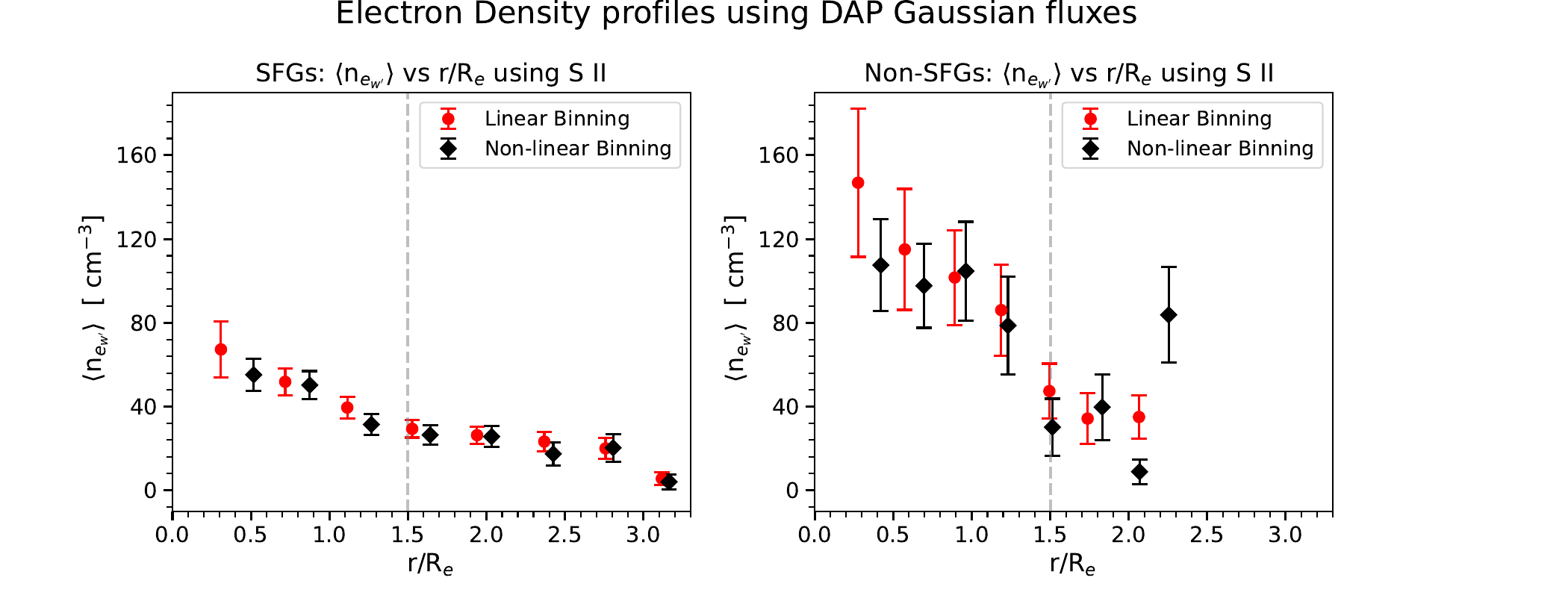}
    \caption{{\bf DAP Profiles:-} \textit{Left panel:} The figure shows the radial profiles of $n_e$(S\,\textsc{ii}) for SFGs. The data points represent the $\langle {n}_{e_{w^{\prime}}} \rangle$ obtained by binning the scatter plots in Fig.~\ref{fig:figure_2} using the volume weight of each galaxy. Red dots and black diamonds depict the linear and non-linear binning cases, respectively. \textit{Right panel:} Similarly, the right panel illustrate the corresponding radial profiles for Non-SFGs. The grey vertical lines in each subplot represent the demarcation at r/R$_e$=1.5. [\textit{Note:} The linear and non-linear binning has been implemented on the $n_e$ maps of each galaxy, not on the scatter plots in Fig.~\ref{fig:figure_2}.]} 
    \label{fig:figure_4}
\end{figure*}

\subsection{Methodology}
\label{subsec:analysis}

\begin{figure*}
    \centering
    \includegraphics[scale=0.5]{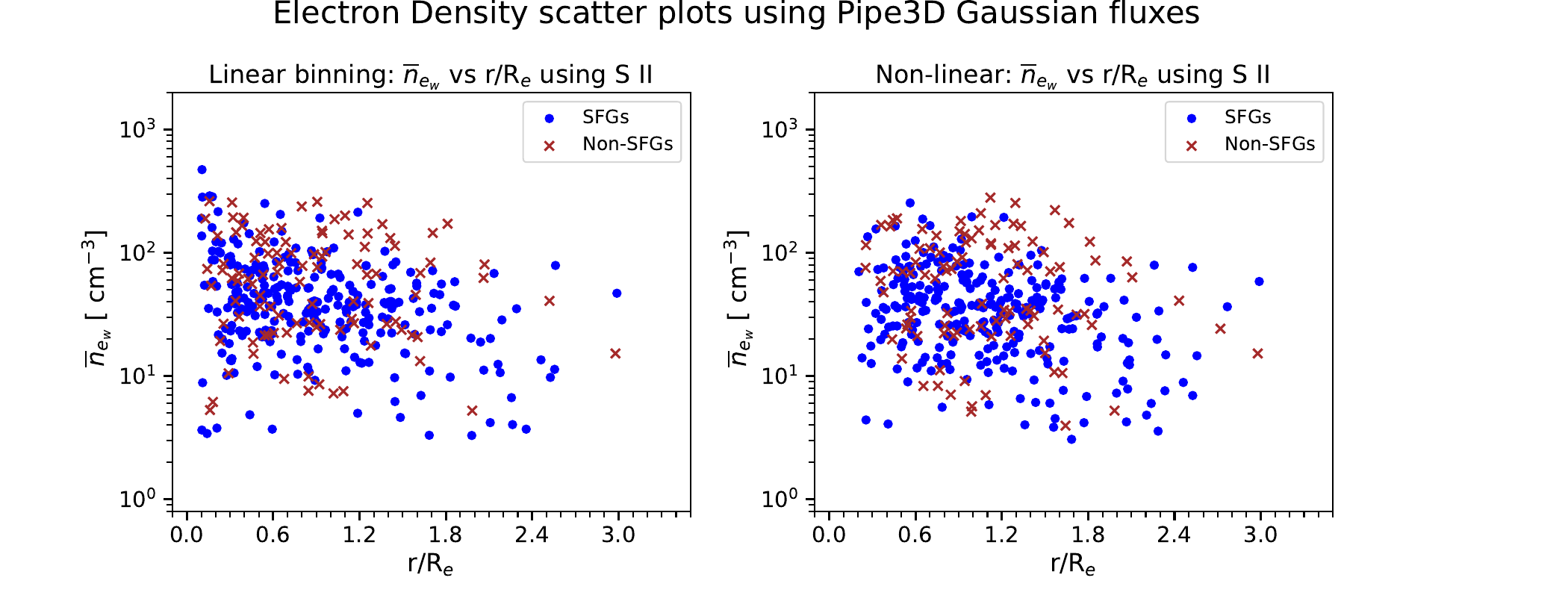}
    \caption{\textit{Left Panel:} The figure illustrates the results from linear binning applied to individual galaxies, using \texttt{Pipe3D} data. The left panel depict scatter plots of $n_e$(S\,\textsc{ii}) for SFGs (blue dots) and Non-SFGs (brown crosses). Each data point corresponds to the $\bar{n}_{e_w}$ computed across different bins within individual galaxies. \textit{Right Panel:} Similarly, the right panel represent the scatter plots for the case of non-linear binning.}
    \label{fig:figure_5}
\end{figure*}

\subsubsection{Flux analysis}
\label{subsubsec:flux_analysis}
In this analysis, we have used the emission line fluxes from the MaNGA DAP and \texttt{Pipe3D}. Both pipelines provide the line fluxes obtained by fitting a Gaussian to the emission lines and using weighted moment analysis. In this work, we have utilised the Gaussian-fitted fluxes of the S\,\textsc{ii} doublets from both DAP and \texttt{Pipe3D}.

{\bf DAP analysis:} The S\,\textsc{ii} line fluxes for each galaxy have been taken from the data products obtained from \cite{Westfall_2019} and \cite{Belfiore_2019} provided in the extension ``EMLINE\_GFLUX”. We begin the analysis by plotting the spatially resolved maps of the fluxes s$_1$ and s$_2$. We find that the outer regions of some of the galaxies show certain pixels having negative flux values, which arise due to incorrect sky calibration and subtraction. For each galaxy, we have masked those pixels and divided each spatial map into annular bins of equal width. Only those bins were chosen for further analysis where the masked pixels were less than 30\%. This condition allowed us to remove 1 SFG and 53 Non-SFGs since the spatial coverage in the remaining bins of these galaxies was not significant enough to carry out further analysis. Hence, this reduces our sample to 66 galaxies with 46 SFGs and 20 Non-SFGs.
In the 66 galaxies, the masked pixels were interpolated in the resulting maps using bilinear interpolation. We have used the convolution and interpolation functions from Astropy\footnote{\href{https://docs.astropy.org/en/latest/convolution/index.html}{www.astropy.org/interpolation}}, where these methods are well implemented.

\begin{figure*}
    \centering
    \includegraphics[scale=0.5]{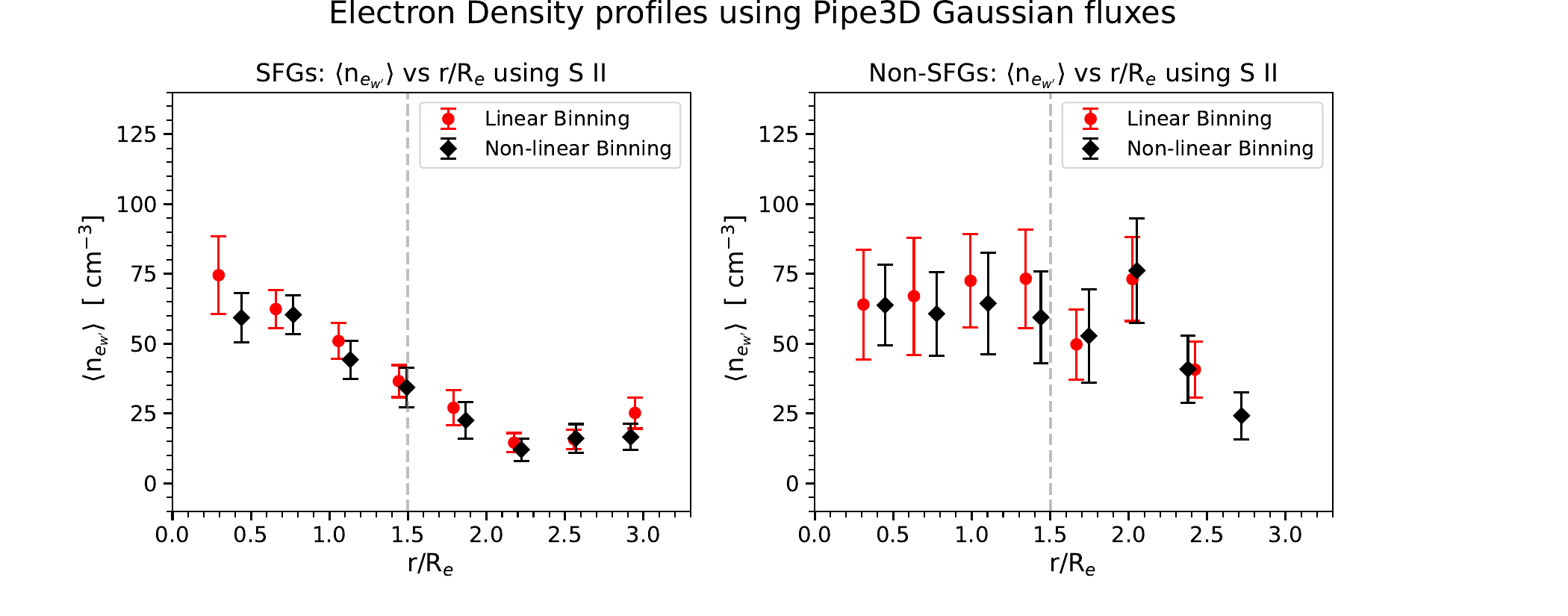}
    \caption{{\bf \texttt{Pipe3D} Profiles:-} \textit{Left panel:} The figure shows the radial profiles of $n_e$(S\,\textsc{ii}) for SFGs. The data points represent the $\langle {n}_{e_{w^{\prime}}} \rangle$ obtained by binning the scatter plots in Fig.~\ref{fig:figure_5} using the volume weight of each galaxy. Red dots and black diamonds depict the linear and non-linear binning cases, respectively. \textit{Right panel:} Similarly, the right panel illustrate the corresponding radial profiles for Non-SFGs. The grey vertical lines in each subplot represent the demarcation at r/R$_e$=1.5. [\textit{Note:} The linear and non-linear binning has been implemented on the $n_e$ maps of each galaxy, not on the scatter plots in Fig.~\ref{fig:figure_5}].}
    \label{fig:figure_7}
\end{figure*}

\begin{figure*}
    \centering
    \includegraphics[scale=0.5]{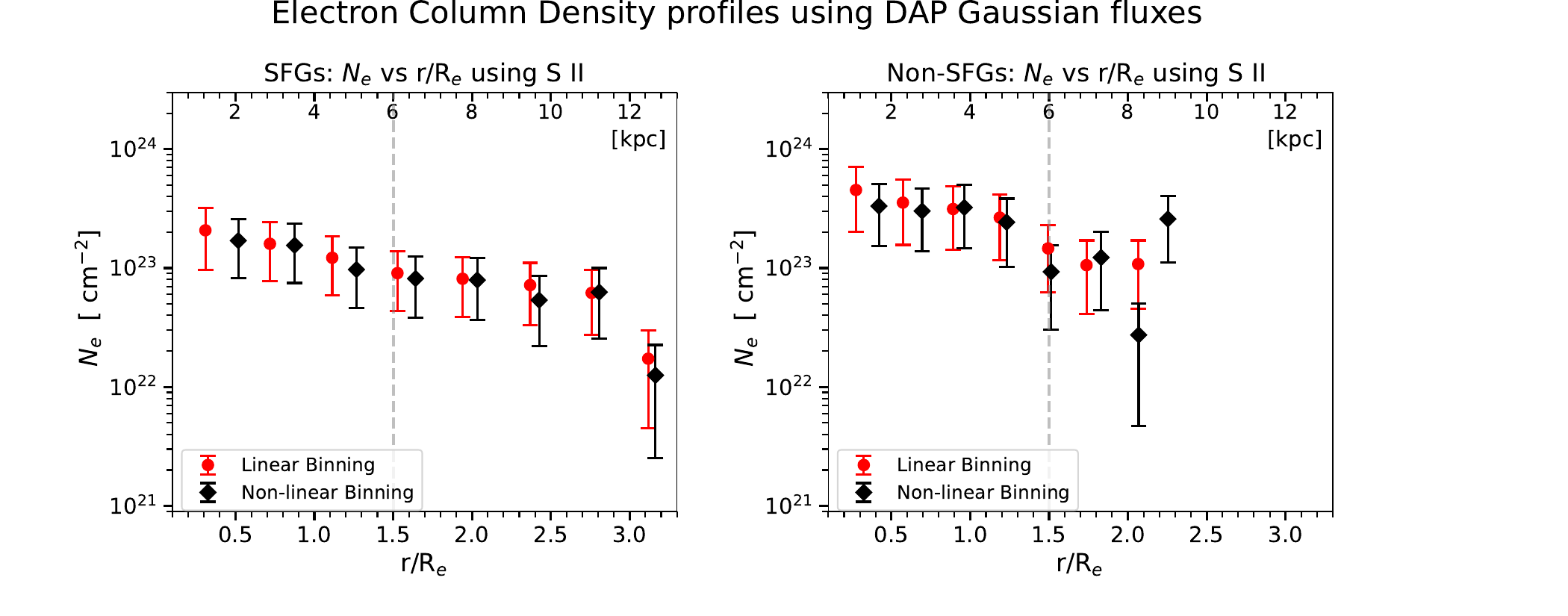}
    \caption{The figure depicts the electron column density profiles obtained using DAP S\,\textsc{ii} fluxes. The top x-axis shows the physical distance in kpc at the median redshift of $z_{med}=0.0371$ for the SFGs sample (and $z_{med}=0.0506$ for Non-SFGs in the right panel).The error on the electron column density is derived from uncertainties in $n_e$ as well as path length}. The left and right panels illustrate the case of $N_e$(S\,\textsc{ii}) for SFGs and Non-SFGs, respectively, derived from the corresponding $n_e$ profiles in Fig.~\ref{fig:figure_4}. The linear and non-linear cases are shown by red-dotted and black diamonds, respectively.
    \label{fig:figure_8}
\end{figure*}

The flux maps were translated to the flux ratio R$_{\text{S\,\textsc{ii}}}$. As discussed in subsection~\ref{subsec:elec_den_theory}, the dependencies of electron density on R are sensitive to a range of R values. Therefore, we choose the R$_{\text{S\,\textsc{ii}}}$ values to lie within the range of $0.4315- 1.449$ (excluding 0.4315), which results in patchiness in the spatial maps of R$_{\text{S\,\textsc{ii}}}$. A careful analysis of the histograms of R$_{\text{S\,\textsc{ii}}}$ showed that the distributions are skewed and highly affected by outliers. To remove the outliers, we find the lower and upper bounds for R$_{\text{S\,\textsc{ii}}}$ distributions using the interquartile range (IQR)\footnote{${\rm IQR} := Q_3 - Q_1$ where $Q_1$ and $Q_3$ are the first and the third quantiles, respectively. The data points below the lower, $L = Q_1 - 1.5 \cdot {\rm IQR}$, and above the upper, $U = Q_3 + 1.5 \cdot {\rm IQR}$, thresholds are rejected as outliers. This method is very robust for skewed distributions.} outlier removal technique. 
The clipped pixels having R$_{\text{S\,\textsc{ii}}}$ $\leq$ 0.4315 and R$_{\text{S\,\textsc{ii}}}$ $>$ 1.449 were replaced by the lower bound and upper bound, respectively. By using Eq.~\ref{eqn:equation_ne}, we calculate the electron densities, $n_e$(S\,\textsc{ii}) from R$_{\text{S\,\textsc{ii}}}$.

{\bf \texttt{Pipe3D} analysis:} 
In this case, we used the Gaussian fitted emission lines fluxes of S\,\textsc{ii} doublet, provided in the ``$\text{ELINES}$" extension. The rest of the analysis follows in the same way as the DAP analysis and electron density maps were obtained for \texttt{Pipe3D} data. Due to the unavailability of uncertainties on the \texttt{Pipe3D} Gaussian fluxes, we have assigned the error from the \texttt{Pipe3D} moment analysis fluxes to the \texttt{Pipe3D} Gaussian fluxes at each pixel. In addition to that, we have also performed the analysis by assigning the uncertainties from the DAP Gaussian fluxes (see Appendix~\ref{appendix:appendix_2}).

\subsubsection{Radial profiles}
\label{subsubsec:radial_profiles}

The analysis of an individual galaxy is presented in Fig.~\ref{fig:single_galaxy}. The left and middle panels display the interpolated DAP flux maps for s$_1$ (upper left), s$_2$ (upper middle), o$_1$ (lower left), and o$_2$ (lower middle) for the galaxy ``1-606221". To obtain the radial profile of electron density, we have implemented two types of angularly averaged radial binning schemes on the electron density maps which are described below: (i) linear binning: in this binning scheme the consecutive bins are equally spaced, (ii) non-linear binning: in this binning scheme all the bins contain almost equal number of pixels (with an additional criterion that the outermost bin must have at least one-pixel radial separation). Among each bin, the weighted average of the electron densities has been calculated as,

\begin{equation}
    \bar{n}_{e_w} = \sum\limits_{i} {w_i}  {n_{e,i}}\Big/\sum\limits_{i} w_i
    \label{eqn:equation_wtd_avg}
\end{equation}

where $w_i = 1/\delta n_{e,i} ^2$ is the weight and $\delta n_{e,i}$ is the uncertainty in $n_e$ on the $i^{th}$ pixel. The error on the weighted mean is calculated as the standard error with the standard deviation calculated from the unbiased weighted variance. The yellow annuli in Fig.~\ref{fig:single_galaxy} show an example of linear binning.

Fig.~\ref{fig:figure_2} illustrates the radial distributions of the electron density for individual galaxies among SFGs (blue dots) and Non-SFGs (brown crosses), respectively, employing linear and non-linear binning in the left and right panels, respectively, for the DAP data. Each data point represents the weighted average of the electron density calculated using Eq.~\ref{eqn:equation_wtd_avg} in different bins of each galaxy. We find that the individual galaxy profiles do not differ much when comparing the two binning schemes. Also, we observe that the scatter plots of $\bar{n}_{e_w}$ exhibit a qualitative decreasing trend in the case of both SFGs and Non-SFGs. To quantify the effect statistically, we distribute the data points into bins of radial distance r/R$_e$. Since the MaNGA sample is not volume-limited, any statistical analysis requires a volume correction for all the galaxies, as discussed in detail by \cite{Wake_2017}. Our sample includes galaxies from ``primary+" and ``secondary" MaNGA subsamples, we adopt the volume weights ($w^{\prime}$) calculated by \cite{Wake_2017}, and provided in the ``\texttt{DAPall}" file by \cite{Belfiore_2019} and \cite{Westfall_2019}.  Also, as discussed in Section~6.4 of \cite{Wake_2017}, the inclination selection can introduce a bias due to higher extinction in edge-on galaxies than in face-on galaxies. Since our sample purely consists of face-on galaxies, there will be no such bias in our analysis and the standard volume weights can be used. Thus, we calculate the weighted average of electron densities in each radial bin as,
\begin{equation}
    \langle {n}_{e_{w^{\prime}}} \rangle = \sum\limits_{j} {w_j^{\prime}}  \bar{n}_{e_w}\Big/\sum\limits_{j} w_j^{\prime}
    \label{eqn:wtd_density_volume_wts}
\end{equation}
and the error on this averaged density is calculated using error propagation as,

\begin{equation}
    \delta\langle {n}_{e_{w^{\prime}}} \rangle = \sum\limits_{j} {w_j^{\prime}}  \delta\bar{n}_{e_w}\Big/\sum\limits_{j} w_j^{\prime}
    \label{eqn:error_wtd_avg_vol_wts}
\end{equation}
where the index `$j$' represents any data point from a particular galaxy in the scatter plots (Fig.~\ref{fig:figure_2}). We also carried out a similar analysis using the \texttt{Pipe3D} data. Fig.~\ref{fig:figure_5} is described in the same manner as Fig.~\ref{fig:figure_2}. The obtained profiles have been discussed in detail in subsection~\ref{subsec:results}.

\subsubsection{Electron Column Density}
\label{subsubsec:column_density}

Noting that the observed line emission is always an integrated contribution along the line of sight, the more useful physical measure is the electron column density, $N_e := \mathop{\int} n_e \, dl$, a line integral through the length of the medium. 
The absorption line studies have been used as a successful tool to obtain the gas column densities (see \cite{Werk_2014} and \cite{Tumlinson_2017}). 
Translating the $n_e$ to $N_e$ also depends on the galactic disk model to be considered. We assume an evenly distributed thin disk of a constant thickness of 1 kpc with 50\% uncertainty for the region outside the central bulge. The disk thickness may be influenced by flaring and vertical gradients; however, this would not significantly affect the overall order of $N_e$, although it could alter the precise values of the estimates. We also use a second approach to estimate the thickness of the emitting line of sight and electron column density discussed in Appendix~\ref{appendix:appendix_path_length}.

To obtain the column density profiles, we have translated the profiles from Fig~\ref{fig:figure_4} using 1 kpc disk thickness. The uncertainties in $N_e$ are calculated from uncertainties in $n_e$ as well as path length, by using error propagation. The profiles are depicted in Fig.~\ref{fig:figure_8} and will be discussed shortly.

\subsection{Results}
\label{subsec:results}
In this subsection, we discuss the outcomes of the analysis done so far. Fig.~\ref{fig:figure_4} shows the radial profiles of volume-weighted average electron density, $\langle n_{e_{w^{\prime}}} \rangle$, for the DAP case. The vertical grey dashed lines show a demarcation at r/R$_e$ = 1.5 to segregate the inner and outer disk regions. The average densities are estimated below and above this demarcation for each profile that has been discussed in this subsection below. The values of all the average densities are provided in Table~\ref{table:table_2}. The left panel depict the $n_e$(S\,\textsc{ii}) profiles for SFGs. The red dots and black diamonds represent the linear and non-linear binning cases, respectively, and the same convention has been followed throughout this paper. We find that the electron density decreases with increasing radial distance. The average density at r/R$_e$ $\leq$ 1.5 and r/R$_e$ $>$ 1.5, are found to be $52.87\pm8.32$ $\text{cm}^{-3}$ and $20.92\pm4.2$ $\text{cm}^{-3}$, respectively, for the linear binning case, and $45.63\pm6.36$ $\text{cm}^{-3}$ and $18.78\pm5.05$ $\text{cm}^{-3}$, respectively, for non-linear binning. Hence, the gradients in both profiles are clearly evident by the average densities. These averages also indicate that both binning schemes are consistent with each other, which can be seen in the profiles. Similarly, the lower right panel represent the $n_e$(S\,\textsc{ii}) profiles for Non-SFGs. We observe similar trends in $n_e$(S\,\textsc{ii}) with radial distance. The average $n_e$(S\,\textsc{ii}) below and above the demarcation, with linear binning, are $99.39 \pm 24.37$ $\text{cm}^{-3}$ and $34.64 \pm 11.24$ $\text{cm}^{-3}$, respectively and those with non-linear binning are $97.13 \pm 22.22$ $\text{cm}^{-3}$ and $40.64 \pm 14.5$ $\text{cm}^{-3}$, respectively.

In the case of \texttt{Pipe3D}, we have employed a similar procedure that was implemented in the DAP case to obtain the radial profiles from the scatter plots (Fig.~\ref{fig:figure_5}). Fig.~\ref{fig:figure_7} illustrates the variation of $\langle n_{e_{w^{\prime}}} \rangle$ with r/R$_e$. For SFGs, in the left panel, we observe decreasing trends of $n_e$(S\,\textsc{ii}) with increasing radial distance. The $n_e$(S\,\textsc{ii}) averages below and above the demarcation are $56.13 \pm 8.25$ $\text{cm}^{-3}$ and $20.65 \pm 4.71$ $\text{cm}^{-3}$, respectively, for linear binning and $49.58 \pm 7.39$ $\text{cm}^{-3}$ and $16.83 \pm 5.11$ $\text{cm}^{-3}$, respectively, for non-linear binning. It is interesting to note that in SFGs, both DAP and \texttt{Pipe3D} analysis yield radial profiles that exhibit similar trends in $n_e$(S\,\textsc{ii}). However, for Non-SFGs, distinct variations are observed. The lower left panel illustrates the radial variation of $n_e$(S\,\textsc{ii}), where the electron density gets flattened at around $\sim$60 cm$^{-3}$ for both binning cases. 

\begin{table*}
\centering
\begin{center}
\resizebox{1.0\textwidth}{!}{ 
\begin{tabular}{|l|c|c|c|c|c|c|c|c|} 
\hline 
& \multicolumn{4}{c|}{SFGs} & \multicolumn{4}{c|}{Non-SFGs} \\ \hline
 & \multicolumn{2}{c|} {r/R$_e$ $\leq 1.5$}  & \multicolumn{2}{c|}{r/R$_e$$> 1.5$ } & \multicolumn{2}{c|}{r/R$_e$ $\leq 1.5$ } & \multicolumn{2}{c|}{r/R$_e$$> 1.5$ } \\ \hline
& Linear & Non-linear & Linear & Non-linear & Linear & Non-linear & Linear & Non-linear \\ \hline

DAP (S~\textsc{ii}) & 52.87 $\pm$ 8.32 & 45.63 $\pm$ 6.36 & 20.92 $\pm$ 4.2 & 18.78 $\pm$ 5.05 & 99.39 $\pm$ 24.37 & 97.13 $\pm$ 22.22 & 34.64 $\pm$ 11.24 & 40.64 $\pm$ 14.5 \\

Pipe3D (S~\textsc{ii}) & 56.13 $\pm$ 8.25 & 49.58 $\pm$ 7.39 & 20.65 $\pm$ 4.71 & 16.83 $\pm$ 5.11 & 69.22 $\pm$ 18.74 & 62.11 $\pm$ 16.01 & 54.54 $\pm$ 12.56 & 48.53 $\pm$ 13.99  \\ \hline 

\end{tabular}
}
\caption{In this table, we present the average electron densities corresponding to the radial profiles depicted in Fig.~\ref{fig:figure_4} (DAP) and Fig.~\ref{fig:figure_7} (\texttt{Pipe3D}). For each case, the electron densities are averaged over two regions: r/R$_e$ $\leq$ 1.5  and r/R$_e$ $>$ 1.5, for both binning schemes. All the presented values are in $\text{cm}^{-3}$.}
\label{table:table_2}
\end{center}
\end{table*}

\section{Discussion}
\label{sec:discussion}

Over the past two decades, advancements in high spectral resolution instruments have made the study of electron density in galaxies increasingly prevalent. In the existing literature, the [S\,\textsc{ii}] and [O\,\textsc{ii}] doublets are the most common tools that have been used to estimate the electron densities. There have been several studies, for example, \cite{Sanders_2016,Kaasinen_2017,Davies_2021}, \cite{Herrera-Camus_2016,Isobe_2023} have probed $n_e$ at different redshift using various lines and surveys. While these studies offer valuable insights into the redshift evolution of electron density, they primarily rely on integrated measurements that average over the entire galaxy, thereby lacking information on spatial variations in the galactic disk. This limitation becomes particularly important in the context of magnetic field studies that utilise Faraday Rotation, where the observed RM is influenced by the local electron density at the specific impact parameter along the quasar sightline. Therefore, the distribution of $n_e$ within the ISM and CGM of galaxies remains questionable, which we have addressed in this work.

\begin{figure*}
    \centering
    \includegraphics[scale=0.5]{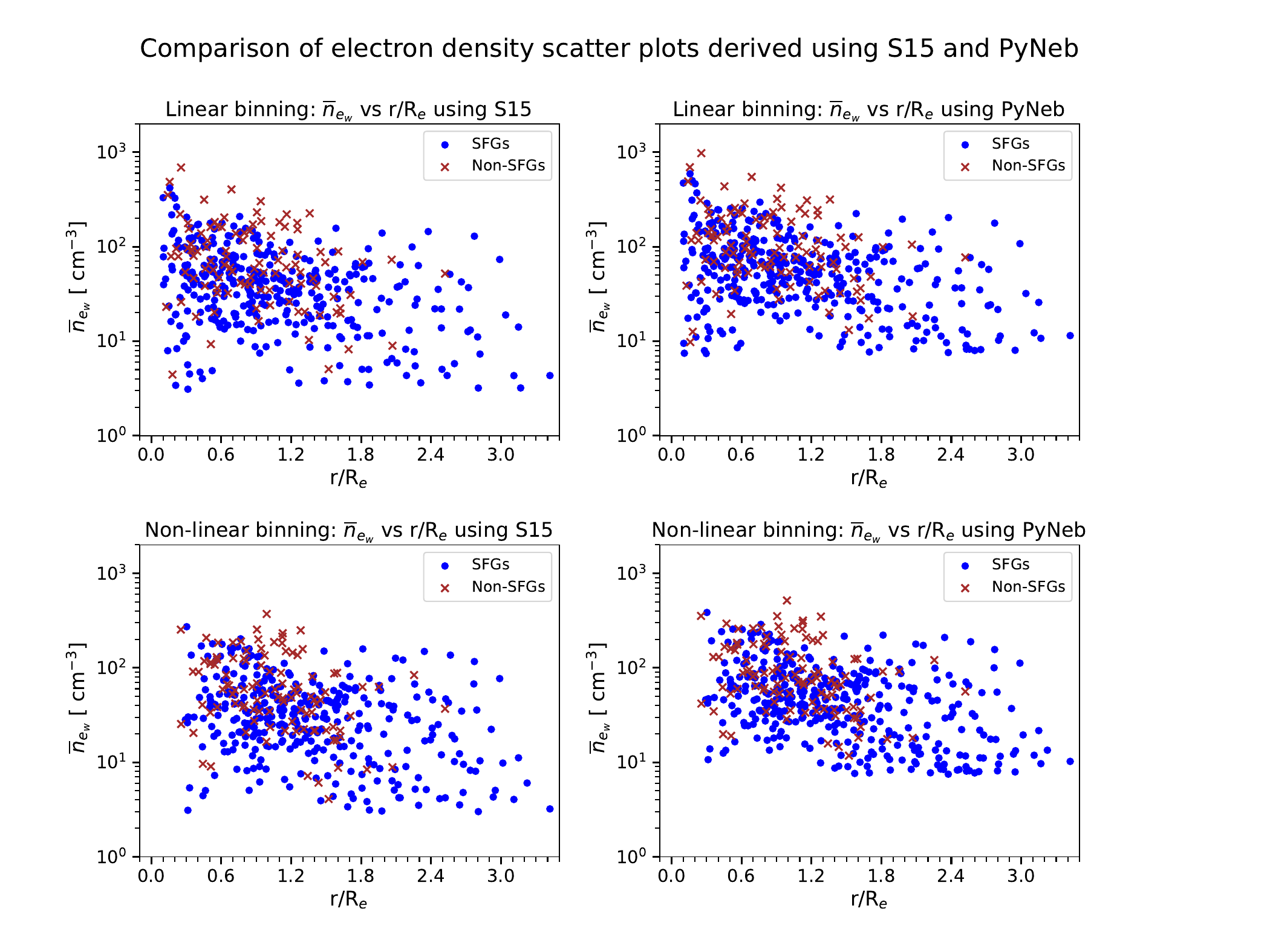}
    \caption{\textit{Top Panel:} The figure illustrates the results from linear binning applied to individual galaxies, using DAP data. The left panel depicts scatter plots of $n_e$(S\,\textsc{ii}) for SFGs (blue dots) and Non-SFGs (brown crosses), computed using the S15 prescription. Each data point corresponds to the $\bar{n}_{e_w}$ calculated across different bins within individual galaxies. \textit{Right Panel:} Similarly, the right panel represent the corresponding scatter plots for the case of non-linear binning, with electron density derived using PyNeb.}
    \label{fig:figure_pyneb_scatter}
\end{figure*}

\begin{figure*}
    \centering
    \includegraphics[scale=0.5]{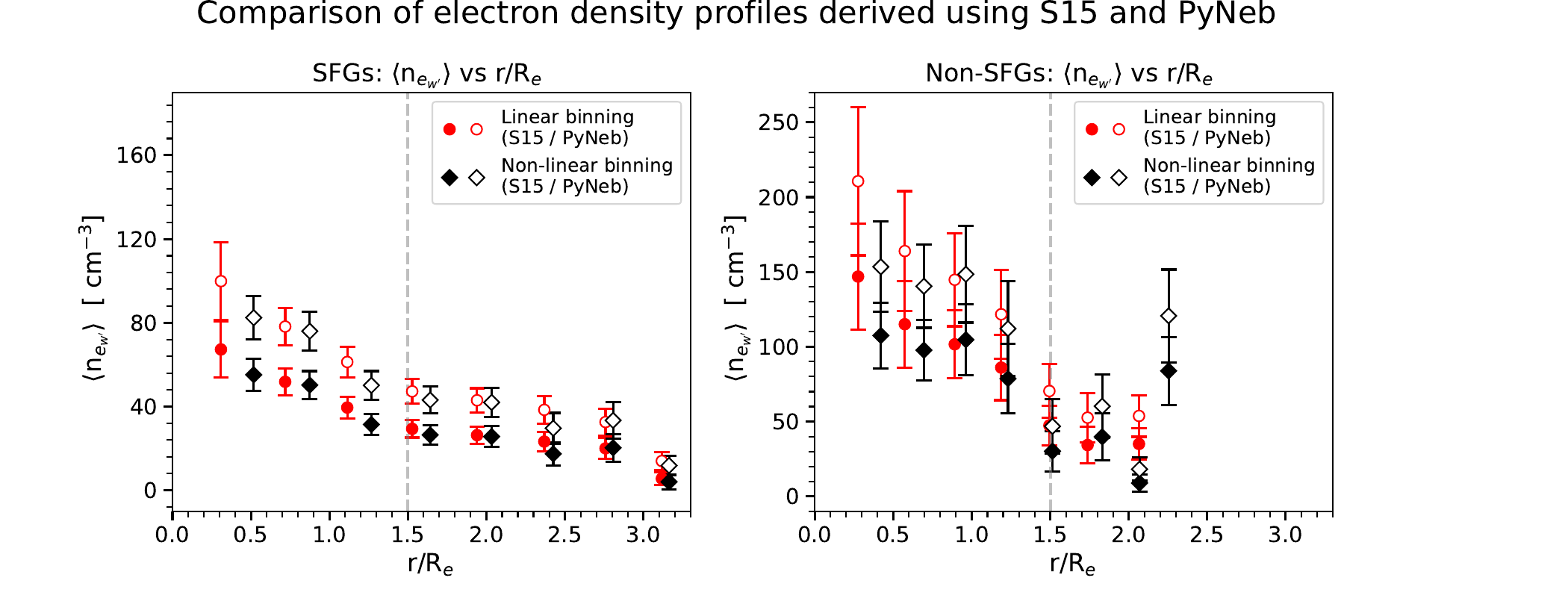}
    \caption{\textit{Left panel:} Figure shows the comparison of the radial profiles of $\langle {n}_{e_{w^{\prime}}} \rangle$, similar to Figure~\ref{fig:figure_4}, obtained by using the S15 prescription (filled circles/diamonds) and PyNeb (open circles/diamonds). Red dots and black diamonds corresponds to the linear and non-linear binning cases, respectively. \textit{Right panel:} Similarly, the right panel illustrate the corresponding radial profiles for Non-SFGs. The grey vertical lines in each subplot represent the demarcation at r/R$_e$=1.5.} 
    \label{fig:figure_pyneb_profiles}
\end{figure*}

We have presented the radial profiles of thermal electron densities in the ISM of nearby face-on galaxies ($z_{med} \sim 0.0405$). The electron densities are calculated using collisionally excited/de-excited [S\,\textsc{ii}] $\lambda\lambda$6716, 6731 \AA~(\cite{Kewley_2019}). The electron density profiles obtained using [O\,\textsc{ii}] $\lambda\lambda$3726, 3729 \AA~doublet are also discussed in Appendix~\ref{appendix:appendix_oii}. We have categorised our sample into SFGs and Non-SFGs based on sSFR. The analysis was done using data from two pipelines: MaNGA DAP and \texttt{Pipe3D}, including two different binning schemes on the $n_e$ maps of each galaxy. 
We find that in star-forming galaxies, both the DAP and \texttt{Pipe3D} $n_e$(S\,\textsc{ii}) profiles are consistent with each other. It is found that the galaxies have higher densities in the inner disk region than the outskirts, for both DAP and \texttt{Pipe3D} cases. On the other hand, in Non-SFGs, we find discrepancies in the estimates from both the DAP and \texttt{Pipe3D} pipelines' profiles.Hence, the electron density analysis for Non-SFGs requires further investigation and will be explored in the future work. In addition to the S15 prescription of \cite{Sanders_2016}, we also derive electron densities using PyNeb (\cite{pyneb_2015}, \cite{Morisset_2020}), which computes level populations by solving the statistical equilibrium equations for multi-level atoms using updated atomic data, to calculate the electron density. Figure~\ref{fig:figure_pyneb_scatter} illustrates the scatter plots of $\bar{n}_{e_w}$ obtained using DAP S\,\textsc{ii} fluxes. The upper left panel shows the linear binning case with $n_e$ calculated using S15 technique. The SFGs and Non-SFGs are represented by blue dots and brown crosses, respectively. The upper right panel depicts the corresponding profiles where the $n_e$ is calculated using PyNeb. Similarly, the lower two panels show the case for non-linear binning. We find that both S15 and PyNeb exhibit similar qualitative trends PyNeb systematically yielding higher electron densities than S15. To illustrate this more clearly, we present the corresponding binned radial profiles in Figure~\ref{fig:figure_pyneb_profiles}, in the same manner as in Figure~\ref{fig:figure_4}. In this figure, the filled circles and diamonds denote the S15 profiles, while the open circles and diamonds represent the PyNeb profiles. We observe that both the approaches exhibit similar radial gradients in the average electron density across the galactic disc.

In addition to the radial profiles, to check the consistency of the observed gradients, we have estimated the average densities in the inner and outer disk regions.
We find that the all the $n_e$(S\,\textsc{ii}) averages obtained using Gaussian fluxes are well constrained within the estimates $\sim$100 $\text{cm}^{-3}$ obtained by BB23 and $\sim$250 $\text{cm}^{-3}$ inferred from the radial profiles provided by EP22. 

There are other studies that probe the radial profiles $n_e$ at the length scales of typical star-forming regions (see \cite{PHILLIPS_2008} to the galactic scales (see EP22 and BB23). EP22 utilised the galaxies from the CALIFA survey and BB23 utilised the entire MaNGA sample and obtained the S\,\textsc{ii} fluxes using \texttt{Pipe3D} pipeline. The authors categorised their samples based on the morphological classifications of the galaxies. Both studies obtained nearly flattened $n_e$(S\,\textsc{ii}) profiles in almost all the cases. Our study provides useful results for various applications. 

For example, several studies, including those by \cite{Bernet2008, bernet_extent_2013,Kronberg_2008,Malik_2020,Burman_2024}, have used a constant electron column density estimate of the order of $10^{20}$ cm$^{-2}$ in their analyses of galactic-scale magnetic fields using Faraday rotation. This assumption plays a critical role in estimating the magnetic field strength along the line of sight. We obtained the column density varies from $\sim$10$^{24}$ cm$^{-2}$ in the inner disk region to $\sim$10$^{22}$ cm$^{-2}$ in the outer disk region. It clearly indicates that at much larger distances ($>$15 kpc) or into the CGM, the column densities can reach as low as $10^{19}- 10^{20}$ cm$^{-2}$.

In our study, we evaluated the validity of this approximation by considering the impact parameter at which the line of sight to the background quasar intersects the foreground galaxy. Specifically, we analysed cases where the impact parameter is greater than 20 kpc. In this regime, our electron density estimates from IFU spectroscopic data (via [S\textsc{ii}] line diagnostics) suggest that the total integrated electron column density along the sightline remains consistent with the assumed $10^{20}$ cm$^{-2}$ range used in previous studies. Thus, our findings confirm that the approximation made in earlier works remains valid in scenarios where quasars probe the CGM and outskirts of galaxies at moderate to high impact parameters. This provides further support for the robustness of these magnetic field estimates derived from Faraday rotation studies.

\section{Conclusions}
\label{sec:conclusion}
In this study, we have investigated the radial distribution of thermal electron densities in the disks of nearby face-on galaxies using the emission line flux from IFU observations from the SDSS MaNGA survey. 
\begin{enumerate}

    \item  We estimated the electron densities and its radial variations for the 66 face-on galaxies (46 SFGs and 20 Non-SFGs) by employing the collisionally originated emission doublet lines of S\,\textsc{ii} fluxes from two MaNGA pipelines MaNGA DAP and \texttt{Pipe3D}. After considering volume-weights of each galaxy, we derived an average radial profile, which shows a decreasing trend. 

    \item For SFGs in the DAP and \texttt{Pipe3D} data, $n_e$(S\,\textsc{ii}) show a decreasing profile, with the estimates being $52.87 \pm 8.32$ and $56.13 \pm 8.25$ cm$^{-3}$, respectively, in the inner disk (r/R$_e$ $\leq$1.5 ) to $20.92 \pm 4.2$ and $20.65 \pm 4.71$ cm$^{-3}$, respectively, in the outer disk (r/R$_e$ $>$1.5). Whereas in Non-SFGs, inconsistencies are observed among the DAP and \texttt{Pipe3D} $n_e$(S\,\textsc{ii}) profiles.

   \item We have translated $n_{e}  \ \rm ({cm}^{-3})$ to electron column densities $N_e ( \ \rm {cm}^{-2})$ by considering a thin disk of thickness 1 kpc beyond the bulge region. We noted that the $N_e$ in the $r 
   \sim 14$ kpc of the outer disk region is of the order of $\sim$10$^{22}$ cm$^{-2}$. If extrapolated, this decreasing trend indicates that it would be much lower in the CGM of these galaxies. 
\end{enumerate}

Our electron density estimates at different radii of the galaxies in the MaNGA survey could validate the approximations from past studies and provide valuable observable for the current and future studies of magnetic fields of galaxies in several radio surveys.

\appendix

\section{Electron density profiles comparison: \textit{\texttt{Pipe3D} S\,\textsc{ii} Gaussian fluxes with DAP and Moment errors}}
\label{appendix:appendix_2}

\begin{figure*}
    \centering
    \includegraphics[scale=0.5]{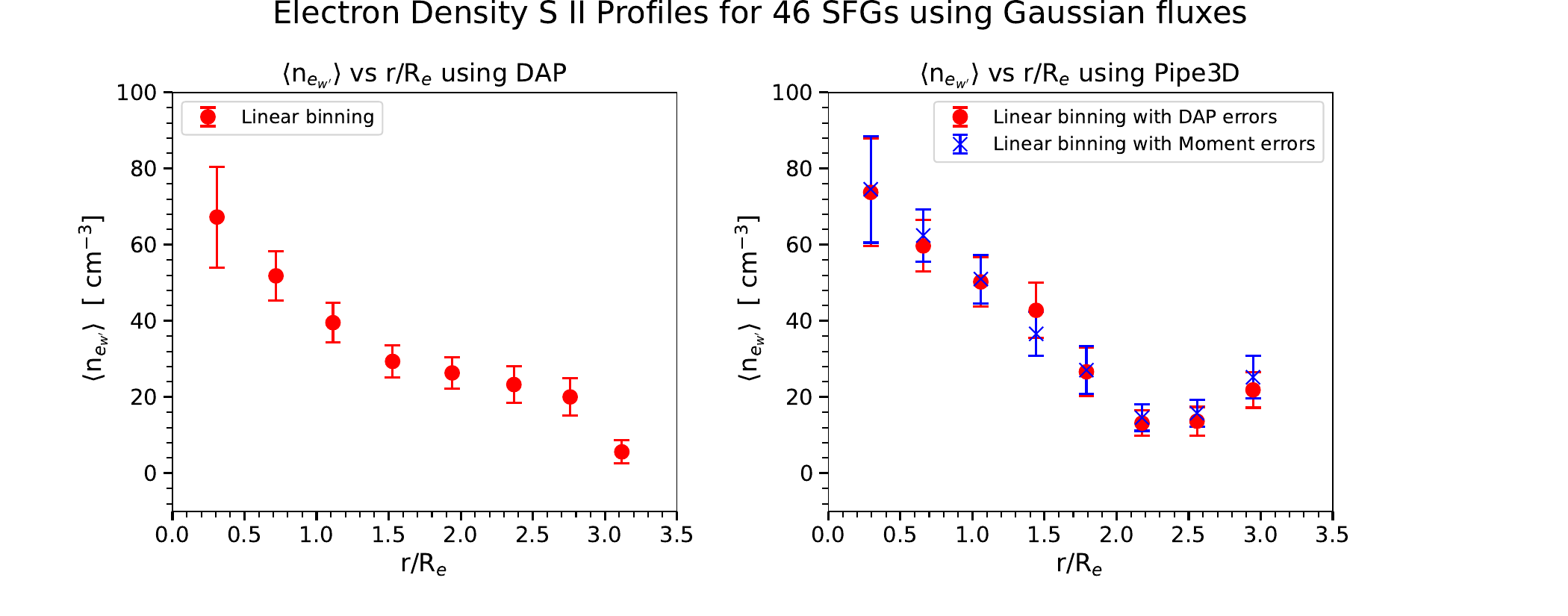}
    \caption{Figure shows the $n_e$(S\,\textsc{ii}) profiles obtained in a similar manner as of Fig.~\ref{fig:figure_4} and Fig.~\ref{fig:figure_7} and only the case of linear binning in individual galaxies. \textit{Left panel:} Electron density profile estimated using DAP Gaussian fluxes. \textit{Right panel:} The profiles are estimated using \texttt{Pipe3D} Gaussian fluxes with DAP errors (red dots) and \texttt{Pipe3D} moment errors (blue crosses).}
    \label{fig:figure_appendix}
\end{figure*}

We have considered two separate cases for S\,\textsc{ii} where we have assigned (i) errors from DAP pipeline to the \texttt{Pipe3D} Gaussian fluxes, and (ii) errors from \texttt{Pipe3D} moment analysis to the \texttt{Pipe3D} Gaussian fluxes. The profiles are depicted in Fig.~\ref{fig:figure_appendix}. We find that the \texttt{Pipe3D} profiles with two different errors assigned agree well with each other.

\section{Electron Density Analysis using [O\,\textsc{ii}] doublet}
\label{appendix:appendix_oii}

We have carried out a similar analysis with [O\,\textsc{ii}] $\lambda\lambda$3726, 3729 \AA~doublet. In this case, we have utilised the Gaussian fluxes from the DAP and the moment analysis fluxes from \texttt{Pipe3D} provided in the extension ``$\text{FLUX}\_\text{ELINES}\_\text{LONG}$” of the datacubes (\cite{Sánchez_2022}). The flux analysis for O\,\textsc{ii} is carried out in the same manner as that for S\,\textsc{ii}. We note that the number of accepted bins in S\,\textsc{ii} and O\,\textsc{ii} differs for some galaxies, which is also illustrated in Figure~\ref{fig:single_galaxy}.

\begin{figure*}
    \centering
    \includegraphics[scale=0.5]{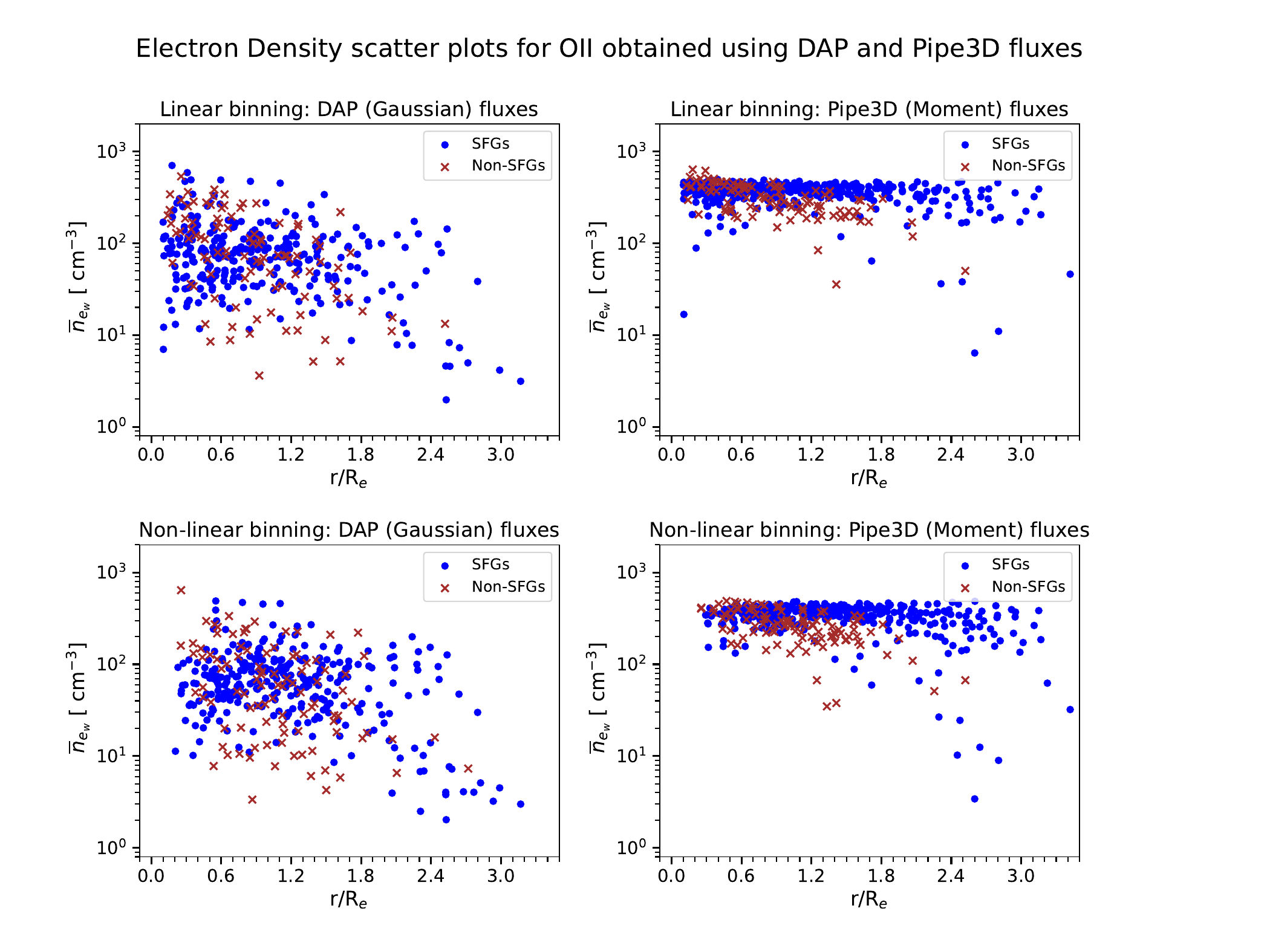}
    \caption{\textit{Upper Panel:} The figure illustrates the results from linear binning applied to individual galaxies, using O\,\textsc{ii} doublet. The left and right panels depict scatter plots obtained using DAP and \texttt{Pipe3D} fluxes, respectively, for SFGs (blue dots) and Non-SFGs (brown crosses). Each data point corresponds to the $\bar{n}_{e_w}$ computed across different bins within individual galaxies. \textit{Lower Panel:} Similarly, the lower panels represent the scatter plots for the case of non-linear binning.}
    \label{fig:figure_appendix_oii_scatter}
\end{figure*}

\begin{figure*}
    \centering
    \includegraphics[scale=0.5]{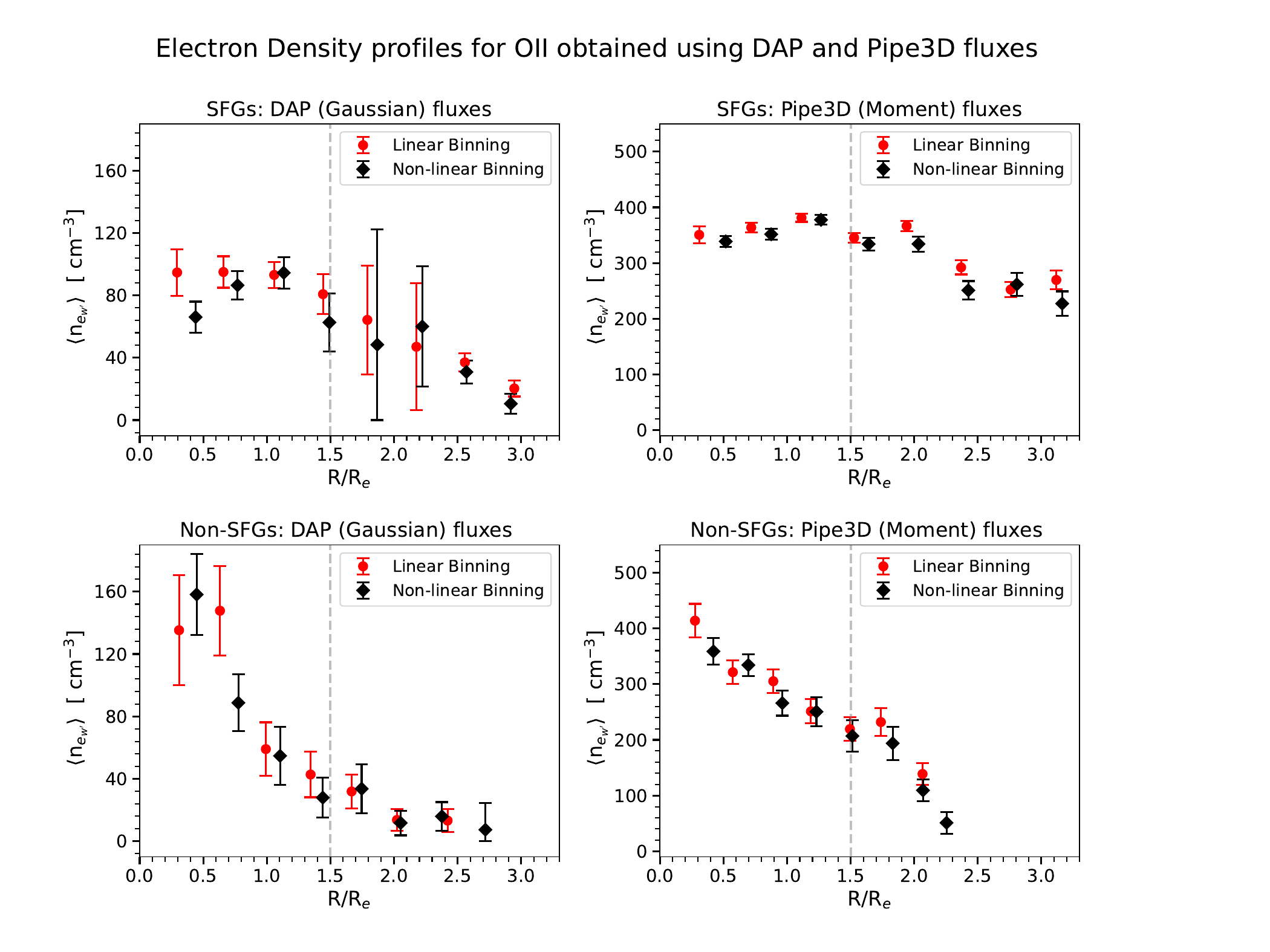}
    \caption{\textit{Upper panel:} The figure shows the radial profiles of $n_e$(O\,\textsc{ii}) obtained using DAP (left) and \texttt{Pipe3D} fluxes (right) for SFGs. The data points represent the $\langle {n}_{e_{w^{\prime}}} \rangle$ obtained by binning the scatter plots in Fig.~\ref{fig:figure_appendix_oii_scatter} using the volume weight of each galaxy. Red dots and black diamonds depict the linear and non-linear binning cases. \textit{Lower panel:} Similarly, the lower panels illustrate the corresponding radial profiles for Non-SFGs. The grey vertical lines in each subplot represent the demarcation at r/R$_e$=1.5. [\textit{Note:} The linear and non-linear binning has been implemented on the $n_e$ maps of each galaxy, not on the scatter plots in Fig.~\ref{fig:figure_appendix_oii_scatter}.]}
    \label{fig:figure_appendix_oii_profiles}
\end{figure*}

The ratios R$_{\text{O\,\textsc{ii}}}$ were calculated from the fluxes. We choose the R$_{\text{O\,\textsc{ii}}}$ values to lie within the range $0.3771-1.457$ (excluding 0.3771). The resulted patchy R$_{\text{O\,\textsc{ii}}}$ maps were translated to $n_e$(O\,\textsc{ii}) and rest of the analysis followed S\,\textsc{ii} as discussed in subsections~\ref{subsubsec:flux_analysis} and \ref{subsubsec:radial_profiles}. Figure~\ref{fig:figure_appendix_oii_scatter} illustrates the radial distributions of the electron density for individual galaxies among SFGs (blue dots) and Non-SFGs (brown crosses), respectively, employing linear and non-linear binning in the top and bottom panels, respectively, for the DAP data in the left panels and \texttt{Pipe3D} data in the right panels.. Each data point represents the weighted average of the electron density, $\bar{n}_{e_w}$, calculated using Eq.~\ref{eqn:equation_wtd_avg} in different bins of each galaxy. In DAP scatter plots, for both binning schemes, a qualitative decreasing trend is observed in the case of both SFGs and Non-SFGs. However, in \texttt{Pipe3D} case no firm trends were seen. The datapoints were radially binned using the volume weights of each galaxy. The resulted profiles are depicted in Figure~\ref{fig:figure_appendix_oii_profiles}. 

The upper panels show the $n_e$(O\,\textsc{ii}) profiles for SFGs with DAP and \texttt{Pipe3D} profiles in the left and right panels, respectively. The linear and non-linear binning are depicted in the similar way as of Figure~\ref{fig:figure_4}. Both the DAP and \texttt{Pipe3D} profiles show decreasing $n_e$(O\,\textsc{ii}) with increasing radial distance; however, the \texttt{Pipe3D} trends are less steeper than the DAP profiles. For the DAP case, the average densities below and above the demarcation are found to be $90.85 \pm 11.58$ $\text{cm}^{-3}$ and $42.14\pm21.61$ $\text{cm}^{-3}$, respectively, for the linear binning case, and $77.38 \pm 11.9$ $\text{cm}^{-3}$ and $37.44 \pm 21.51$ $\text{cm}^{-3}$, respectively, for non-linear binning. Whereas for the \texttt{Pipe3D} case, the $n_e$(O\,\textsc{ii}) averages below and above the demarcation are $365.22 \pm 10.56$ $\text{cm}^{-3}$ and $305.19 \pm 12.25$ $\text{cm}^{-3}$, respectively, for linear binning and $355.95 \pm 9.35$ $\text{cm}^{-3}$ and $281.6 \pm 16.89$ $\text{cm}^{-3}$, respectively, for non-linear binning.

For Non-SFGs, in case of DAP, the $n_e$(O\,\textsc{ii}) averages below and above the demarcation, with linear binning, are $96.22 \pm 23.89$ $\text{cm}^{-3}$ and $19.6 \pm 8.46$ $\text{cm}^{-3}$, respectively and those with non-linear binning are $82.39 \pm 18.91$ $\text{cm}^{-3}$ and $17.14 \pm 12.52$ $\text{cm}^{-3}$, respectively. In the \texttt{Pipe3D} case, the average densities for linear binning are $302.17 \pm 23.03$ $\text{cm}^{-3}$ below the demarcation and $185.29 \pm 22.08$ $\text{cm}^{-3}$ above it, while for non-linear binning, these are $302.31 \pm 23.04$ $\text{cm}^{-3}$ and $140.22 \pm 24.17$ $\text{cm}^{-3}$, respectively. 

We note that the two binning schemes are consistent in the corresponding cases, however there are discrepancies in the magnitude of the densities across the pipelines. This could possibly be due to the choice of Gaussian and moment fluxes from the DAP and \texttt{Pipe3D}, respectively. It is interesting to note that the $n_e$(O\,\textsc{ii}) averages are relatively higher than $n_e$(S\,\textsc{ii}) in the corresponding cases. This difference is expected, as both the doublets are sensitive to different R ranges, hence different electron densities.

\section{Thickness of emitting regions and N$_e$}
\label{appendix:appendix_path_length}

\begin{figure*}
    \centering
    \includegraphics[scale=0.5]{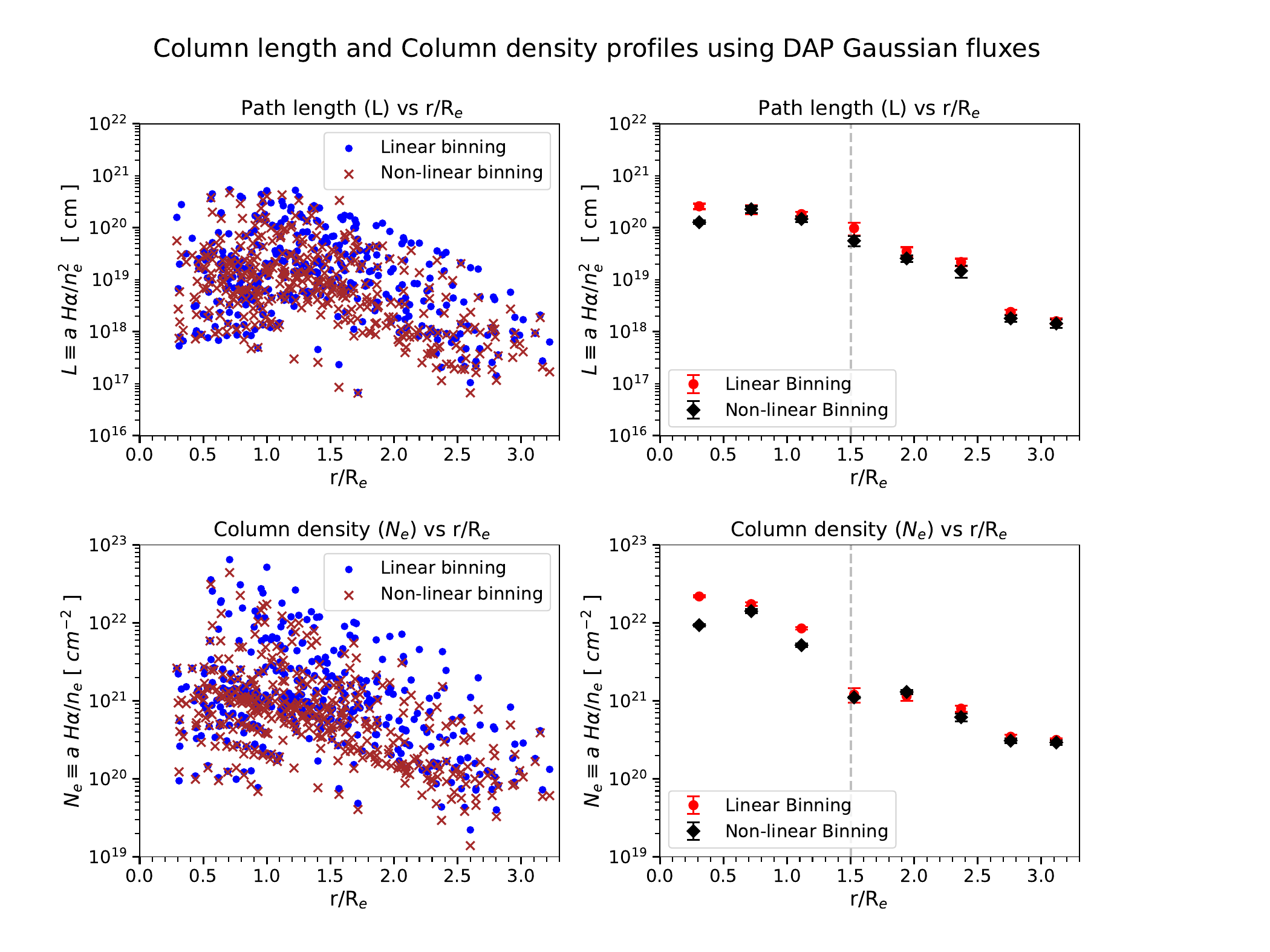}
    \caption{\textit{Upper Panel:} In the left panel, the figure illustrates the results from linear (blue dots) and non-linear (brown crosses) binning applied to individual galaxies, using DAP data. Each data point corresponds to the thickness of the emitting region (L), computed across different bins in each galaxy. The right panel represents the profiles obtained by radially binning the scatter plots and combining the data points using volume weights of each galaxy. \textit{Lower panel:} Similarly, the lower panels depict the scatter plot and profiles of the electron column density ($N_e$) in the left and right panels, respectively. }
    \label{fig:figure_appendix_path_length}
\end{figure*}

We have estimated the thickness of the emitting line of sight, $L$ and electron column density $N_e$ using $H\alpha/n_e^2$ and $H\alpha/n_e$, respectively, for all galaxies, assuming $T=10^4$ K~\citep{Jelic_2015}. The quantities $H\alpha/n_e^2$ and $H\alpha/n_e$ are the proxies to $ff N_e$ and $ff L$, respectively, with $ff = 0.0184~n_e^{-1.07}$ being the gas filling factor (see \cite{Filling_fraction}). We have shown the radial variation in Figure~\ref{fig:figure_appendix_path_length}. We find that these estimates show gradients in both $L$ and $N_e$, with column densities decreasing from $N_e \sim 10^{22}$ cm$^{-2}$ in the inner disc region to $N_e \sim 10^{20}$ cm$^{-2}$ in the outer disc region.

\acknowledgments

We thank the anonymous referees for valuable comments that significantly improved the paper. The research of S.S. is supported by the Core Research Grant CRG/2021/003053 from the Science and Engineering Research Board, India. Research of S.B. is supported by the Institute Fellowship from the Indian Institute of Technology Delhi, India. S.M. and Y.W. acknowledge the support of the Department of Atomic Energy, Government of India, under the project number 12-R\&D-TFR5.02-0700. S.M. also acknowledges support from grant PID2023-146372OB-I00, funded by MICIU/AEI/10.13039/501100011033 and by ERDF, EU

Funding for the Sloan Digital Sky Survey IV has been provided by the Alfred P. Sloan Foundation, the U.S. Department of Energy Ofﬁce of Science, and the Participating Institutions. SDSS-IV acknowledges support and resources from the Center for High Performance Computing at the
University of Utah. The SDSS website is \href{www.sdss.org}{www.sdss.org}. SDSS-IV is managed by the Astrophysical Research Consortium for the Participating Institutions of the SDSS Collaboration including the Brazilian Participation Group, the Carnegie Institution for Science, Carnegie Mellon University, Center for Astrophysics, Harvard \& Smithsonian, the Chilean
Participation Group, the French Participation Group, Instituto de Astrofísica de Canarias, The Johns Hopkins University, Kavli Institute for the Physics and Mathematics of the Universe (IPMU)/University of Tokyo, the Korean Participation Group, Lawrence Berkeley National Laboratory, Leibniz Institut für Astrophysik Potsdam (AIP), Max-Planck-Institut für Astronomie (MPIA Heidelberg), Max-Planck-Institut für Astrophysik (MPA Garching), Max-Planck-Institut für Extraterrestrische Physik (MPE), National Astronomical Observatories of China, New Mexico State University, New York University, University of Notre Dame, Observatário Nacional/MCTI, The
Ohio State University, Pennsylvania State University, Shanghai Astronomical Observatory, United Kingdom Participation Group, Universidad Nacional Autónoma de México, University of Arizona, University of Colorado Boulder, University of Oxford, University of Portsmouth, University of Utah, University of Virginia, University of Washington, University of Wisconsin, Vanderbilt University, and Yale University, and the Collaboration Overview Start Guide Afﬁliate Institutions Key People in SDSS Collaboration Council Committee on Inclusiveness Architects SDSS-IV Survey Science Teams and Working Groups Code of Conduct Publication Policy How to Cite SDSS External Collaborator Policy.

This work made use of Astropy:\footnote{\href{https://www.astropy.org/}{http://www.astropy.org}} a community-developed core Python package and an ecosystem of tools and resources for astronomy \citep{astropy_2013, astropy_2018, astropy_2022}.

\bibliographystyle{JHEP}
\bibliography{references}

\providecommand{\href}[2]{#2}\begingroup\raggedright\begin{thebibliography}{10}

\bibitem{Osterbrock_book_2006}
D.~Osterbrock and G.~Ferland, \emph{Astrophysics Of Gas Nebulae and Active Galactic Nuclei}, G - Reference,Information and Interdisciplinary Subjects Series, University Science Books (2006).

\bibitem{Snell_2019}
R.L.~{Snell} and S.E.~{Kurtz}, \emph{{Fundamentals of Radio Astronomy: Astrophysics (Series in Astronomy and Astrophysics)}} (2019).

\bibitem{Bernet2008}
M.L.~Bernet, F.~Miniati, S.J.~Lilly, P.P.~Kronberg and M.~Dessauges-Zavadsky, \emph{Strong magnetic fields in normal galaxies at high redshift}, \href{https://doi.org/10.1038/nature07105}{\emph{Nature} {\bfseries 454} (2008) 302 – 304}.

\bibitem{bernet_extent_2013}
M.L.~Bernet, F.~Miniati and S.J.~Lilly, \emph{{THE} {EXTENT} {OF} {MAGNETIC} {FIELDS} {AROUND} {GALAXIES} {OUT} {TO} z $\sim$ 1}, \href{https://doi.org/10.1088/2041-8205/772/2/L28}{\emph{ApJL} {\bfseries 772} (2013) L28}.

\bibitem{Malik_2020}
S.~Malik, H.~Chand and T.R.~Seshadri, \emph{Role of intervening mg ii absorbers on the rotation measure and fractional polarization of the background quasars}, \href{https://doi.org/10.3847/1538-4357/ab6bd5}{\emph{The Astrophysical Journal} {\bfseries 890} (2020) 132}.

\bibitem{Burman_2024}
S.~Burman, P.~Sharma, S.~Malik and S.~Singh, \emph{Investigation of the radial profile of galactic magnetic fields using rotation measure of background quasars}, \href{https://doi.org/10.1088/1475-7516/2024/08/063}{\emph{Journal of Cosmology and Astroparticle Physics} {\bfseries 2024} (2024) 063}.

\bibitem{NE2001}
J.M.~{Cordes} and T.J.W.~{Lazio}, \emph{{NE2001.I. A New Model for the Galactic Distribution of Free Electrons and its Fluctuations}}, \href{https://doi.org/10.48550/arXiv.astro-ph/0207156}{\emph{arXiv e-prints} (2002) astro} [\href{https://arxiv.org/abs/astro-ph/0207156}{{\ttfamily astro-ph/0207156}}].

\bibitem{Yao_2017}
J.M.~Yao, R.N.~Manchester and N.~Wang, \emph{A new electron-density model for estimation of pulsar and frb distances}, \href{https://doi.org/10.3847/1538-4357/835/1/29}{\emph{The Astrophysical Journal} {\bfseries 835} (2017) 29}.

\bibitem{Petroff2019}
E.~Petroff, J.W.T.~Hessels and D.R.~Lorimer, \emph{Fast radio bursts}, \href{https://doi.org/10.1007/s00159-019-0116-6}{\emph{The Astronomy and Astrophysics Review} {\bfseries 27} (2019) 4}.

\bibitem{Steidel_2014}
C.C.~Steidel, G.C.~Rudie, A.L.~Strom, M.~Pettini, N.A.~Reddy, A.E.~Shapley et~al., \emph{Strong nebular line ratios in the spectra of z $\sim$ 2–3 star forming galaxies: First results from kbss-mosfire}, \href{https://doi.org/10.1088/0004-637X/795/2/165}{\emph{The Astrophysical Journal} {\bfseries 795} (2014) 165}.

\bibitem{Shimakawa_2015}
R.~{Shimakawa}, T.~{Kodama}, C.C.~{Steidel}, K.-i.~{Tadaki}, I.~{Tanaka}, A.L.~{Strom} et~al., \emph{{Correlation between star formation activity and electron density of ionized gas at z = 2.5}}, \href{https://doi.org/10.1093/mnras/stv915}{\emph{\mnras} {\bfseries 451} (2015) 1284} [\href{https://arxiv.org/abs/1411.1408}{{\ttfamily 1411.1408}}].

\bibitem{Sanders_2016}
R.L.~Sanders, A.E.~Shapley, M.~Kriek, N.A.~Reddy, W.R.~Freeman, A.L.~Coil et~al., \emph{The mosdef survey: Electron density and ionization parameter at z $\sim$ 2.3}, \href{https://doi.org/10.3847/0004-637X/816/1/23}{\emph{The Astrophysical Journal} {\bfseries 816} (2015) 23}.

\bibitem{Davies_2021}
R.L.~Davies, N.M.F.~Schreiber, R.~Genzel, T.T.~Shimizu, R.I.~Davies, A.~Schruba et~al., \emph{The kmos3d survey: Investigating the origin of the elevated electron densities in star-forming galaxies at 1 $\leqslant$ z $\leqslant$ 3}, \href{https://doi.org/10.3847/1538-4357/abd551}{\emph{The Astrophysical Journal} {\bfseries 909} (2021) 78}.

\bibitem{Kaasinen_2017}
M.~Kaasinen, F.~Bian, B.~Groves, L.J.~Kewley and A.~Gupta, \emph{{The COSMOS-[O II] survey: evolution of electron density with star formation rate}}, \href{https://doi.org/10.1093/mnras/stw2827}{\emph{Monthly Notices of the Royal Astronomical Society} {\bfseries 465} (2016) 3220}.

\bibitem{Kashino_2017}
D.~Kashino, J.D.~Silverman, D.~Sanders, J.S.~Kartaltepe, E.~Daddi, A.~Renzini et~al., \emph{The fmos-cosmos survey of star-forming galaxies at z $\sim$ 1.6. iv. excitation state and chemical enrichment of the interstellar medium}, \href{https://doi.org/10.3847/1538-4357/835/1/88}{\emph{The Astrophysical Journal} {\bfseries 835} (2017) 88}.

\bibitem{Harshan_2020}
A.~Harshan, A.~Gupta, K.-V.~Tran, L.Y.~Alcorn, T.~Yuan, G.G.~Kacprzak et~al., \emph{Zfire: Measuring electron density with [o ii] as a function of environment at z = 1.62}, \href{https://doi.org/10.3847/1538-4357/ab76cf}{\emph{The Astrophysical Journal} {\bfseries 892} (2020) 77}.

\bibitem{Herrera-Camus_2016}
R.~Herrera-Camus, A.~Bolatto, J.D.~Smith, B.~Draine, E.~Pellegrini, M.~Wolfire et~al., \emph{The ionized gas in nearby galaxies as traced by the n ii 122 and 205 $\mu$m transitions}, \href{https://doi.org/10.3847/0004-637X/826/2/175}{\emph{The Astrophysical Journal} {\bfseries 826} (2016) 175}.

\bibitem{Kashino_2019}
D.~Kashino and A.K.~Inoue, \emph{Disentangling the physical parameters of gaseous nebulae and galaxies}, \href{https://doi.org/10.1093/mnras/stz881}{\emph{Monthly Notices of the Royal Astronomical Society} {\bfseries 486} (2019) 1053}.

\bibitem{Law_2021}
D.R.~Law, X.~Ji, F.~Belfiore, M.A.~Bershady, M.~Cappellari, K.B.~Westfall et~al., \emph{Sdss-iv manga: Refining strong line diagnostic classifications using spatially resolved gas dynamics}, \href{https://doi.org/10.3847/1538-4357/abfe0a}{\emph{The Astrophysical Journal} {\bfseries 915} (2021) 35}.

\bibitem{Belfiore_2022}
{Belfiore, F.}, {Santoro, F.}, {Groves, B.}, {Schinnerer, E.}, {Kreckel, K.}, {Glover, S. C. O.} et~al., \emph{A tale of two digs: The relative role of hii regions and low-mass hot evolved stars in powering the diffuse ionised gas (dig) in phangs–muse galaxies}, \href{https://doi.org/10.1051/0004-6361/202141859}{\emph{\aap} {\bfseries 659} (2022) A26}.

\bibitem{Espinosa_CALIFA_2022}
C.~Espinosa-Ponce, S.F.~Sánchez, C.~Morisset, J.K.~Barrera-Ballesteros, L.~Galbany, R.~García-Benito et~al., \emph{H ii regions in califa survey: Ii. the relation between their physical properties and galaxy evolution}, \href{https://doi.org/10.1093/mnras/stac456}{\emph{Monthly Notices of the Royal Astronomical Society} {\bfseries 512} (2022) 3436}.

\bibitem{CALIFA_survey_2012}
{Sánchez, S. F.}, {Kennicutt, R. C.}, {Gil de Paz, A.}, {van de Ven, G.}, {Vílchez, J. M.}, {Wisotzki, L.} et~al., \emph{Califa, the calar alto legacy integral field area survey - i. survey presentation}, \href{https://doi.org/10.1051/0004-6361/201117353}{\emph{A\&A} {\bfseries 538} (2012) A8}.

\bibitem{Espinosa_Ponce_2020}
C.~Espinosa-Ponce, S.F.~Sánchez, C.~Morisset, J.K.~Barrera-Ballesteros, L.~Galbany, R.~García-Benito et~al., \emph{H ii regions in the califa survey: I. catalogue presentation}, \href{https://doi.org/10.1093/mnras/staa782}{\emph{Monthly Notices of the Royal Astronomical Society} {\bfseries 494} (2020) 1622}.

\bibitem{Barrera2023sdss}
J.~Barrera-Ballesteros, S.~S{\'a}nchez, C.~Espinosa-Ponce, C.~L{\'o}pez-Cob{\'a}, L.~Carigi, A.Z.~Lugo-Aranda et~al., \emph{Sdss-iv manga: The radial distribution of physical properties within galaxies in the nearby universe}, {\emph{Revista mexicana de astronom{\'\i}a y astrof{\'\i}sica} {\bfseries 59} (2023) 213}.

\bibitem{Bundy_2015}
K.~{Bundy}, M.A.~{Bershady}, D.R.~{Law}, R.~{Yan}, N.~{Drory}, N.~{MacDonald} et~al., \emph{{Overview of the SDSS-IV MaNGA Survey: Mapping nearby Galaxies at Apache Point Observatory}}, \href{https://doi.org/10.1088/0004-637X/798/1/7}{\emph{\apj} {\bfseries 798} (2015) 7} [\href{https://arxiv.org/abs/1412.1482}{{\ttfamily 1412.1482}}].

\bibitem{Yan_2016}
R.~Yan, C.~Tremonti, M.A.~Bershady, D.R.~Law, D.J.~Schlegel, K.~Bundy et~al., \emph{Sdss-iv/manga: Spectrophotometric calibration technique}, \href{https://doi.org/10.3847/0004-6256/151/1/8}{\emph{The Astronomical Journal} {\bfseries 151} (2015) 8}.

\bibitem{Planck_2015}
{Planck Collaboration}, {Ade, P. A. R.}, {Aghanim, N.}, {Arnaud, M.}, {Ashdown, M.}, {Aumont, J.} et~al., \emph{Planck 2015 results - xiii. cosmological parameters}, \href{https://doi.org/10.1051/0004-6361/201525830}{\emph{A\&A} {\bfseries 594} (2016) A13}.

\bibitem{Cappellari_Atlas_2011}
M.~Cappellari, E.~Emsellem, D.~Krajnović, R.M.~McDermid, N.~Scott, G.A.~Verdoes~Kleijn et~al., \emph{The atlas3d project – i. a volume-limited sample of 260 nearby early-type galaxies: science goals and selection criteria}, \href{https://doi.org/10.1111/j.1365-2966.2010.18174.x}{\emph{Monthly Notices of the Royal Astronomical Society} {\bfseries 413} (2011) 813}.

\bibitem{Croom_2012}
S.M.~{Croom}, J.S.~{Lawrence}, J.~{Bland-Hawthorn}, J.J.~{Bryant}, L.~{Fogarty}, S.~{Richards} et~al., \emph{{The Sydney-AAO Multi-object Integral field spectrograph}}, \href{https://doi.org/10.1111/j.1365-2966.2011.20365.x}{\emph{\mnras} {\bfseries 421} (2012) 872} [\href{https://arxiv.org/abs/1112.3367}{{\ttfamily 1112.3367}}].

\bibitem{Brodie_2014}
J.P.~Brodie, A.J.~Romanowsky, J.~Strader, D.A.~Forbes, C.~Foster, Z.G.~Jennings et~al., \emph{The sages legacy unifying globulars and galaxies survey (sluggs): Sample definition, methods, and initial results}, \href{https://doi.org/10.1088/0004-637X/796/1/52}{\emph{The Astrophysical Journal} {\bfseries 796} (2014) 52}.

\bibitem{Ma_2014}
C.-P.~Ma, J.E.~Greene, N.~McConnell, R.~Janish, J.P.~Blakeslee, J.~Thomas et~al., \emph{The massive survey. i. a volume-limited integral-field spectroscopic study of the most massive early-type galaxies within 108 mpc}, \href{https://doi.org/10.1088/0004-637X/795/2/158}{\emph{The Astrophysical Journal} {\bfseries 795} (2014) 158}.

\bibitem{Gunn_2006}
J.E.~{Gunn}, W.A.~{Siegmund}, E.J.~{Mannery}, R.E.~{Owen}, C.L.~{Hull}, R.F.~{Leger} et~al., \emph{{The 2.5 m Telescope of the Sloan Digital Sky Survey}}, \href{https://doi.org/10.1086/500975}{\emph{\aj} {\bfseries 131} (2006) 2332} [\href{https://arxiv.org/abs/astro-ph/0602326}{{\ttfamily astro-ph/0602326}}].

\bibitem{Blanton_2017}
M.R.~Blanton, M.A.~Bershady, B.~Abolfathi, F.D.~Albareti, C.A.~Prieto, A.~Almeida et~al., \emph{Sloan digital sky survey iv: Mapping the milky way, nearby galaxies, and the distant universe}, \href{https://doi.org/10.3847/1538-3881/aa7567}{\emph{The Astronomical Journal} {\bfseries 154} (2017) 28}.

\bibitem{Smee_2013}
S.A.~{Smee}, J.E.~{Gunn}, A.~{Uomoto}, N.~{Roe}, D.~{Schlegel}, C.M.~{Rockosi} et~al., \emph{{The Multi-object, Fiber-fed Spectrographs for the Sloan Digital Sky Survey and the Baryon Oscillation Spectroscopic Survey}}, \href{https://doi.org/10.1088/0004-6256/146/2/32}{\emph{\aj} {\bfseries 146} (2013) 32} [\href{https://arxiv.org/abs/1208.2233}{{\ttfamily 1208.2233}}].

\bibitem{Law_2015}
D.R.~Law, R.~Yan, M.A.~Bershady, K.~Bundy, B.~Cherinka, N.~Drory et~al., \emph{Observing strategy for the sdss-iv/manga ifu galaxy survey}, \href{https://doi.org/10.1088/0004-6256/150/1/19}{\emph{The Astronomical Journal} {\bfseries 150} (2015) 19}.

\bibitem{Wake_2017}
D.A.~Wake, K.~Bundy, A.M.~Diamond-Stanic, R.~Yan, M.R.~Blanton, M.A.~Bershady et~al., \emph{The sdss-iv manga sample: Design, optimization, and usage considerations}, \href{https://doi.org/10.3847/1538-3881/aa7ecc}{\emph{The Astronomical Journal} {\bfseries 154} (2017) 86}.

\bibitem{Westfall_2019}
K.B.~Westfall, M.~Cappellari, M.A.~Bershady, K.~Bundy, F.~Belfiore, X.~Ji et~al., \emph{The data analysis pipeline for the sdss-iv manga ifu galaxy survey: Overview}, \href{https://doi.org/10.3847/1538-3881/ab44a2}{\emph{The Astronomical Journal} {\bfseries 158} (2019) 231}.

\bibitem{Belfiore_2019}
F.~Belfiore, K.B.~Westfall, A.~Schaefer, M.~Cappellari, X.~Ji, M.A.~Bershady et~al., \emph{The data analysis pipeline for the sdss-iv manga ifu galaxy survey: Emission-line modeling}, \href{https://doi.org/10.3847/1538-3881/ab3e4e}{\emph{The Astronomical Journal} {\bfseries 158} (2019) 160}.

\bibitem{LACERDA2022101895}
E.A.~Lacerda, S.~Sánchez, A.~Mejía-Narváez, A.~Camps-Fariña, C.~Espinosa-Ponce, J.~Barrera-Ballesteros et~al., \emph{pyfit3d and pypipe3d — the new version of the integral field spectroscopy data analysis pipeline}, \href{https://doi.org/https://doi.org/10.1016/j.newast.2022.101895}{\emph{New Astronomy} {\bfseries 97} (2022) 101895}.

\bibitem{Sánchez_2022}
S.F.~Sánchez, J.K.~Barrera-Ballesteros, E.~Lacerda, A.~Mejía-Narvaez, A.~Camps-Fariña, G.~Bruzual et~al., \emph{Sdss-iv manga: pypipe3d analysis release for 10,000 galaxies}, \href{https://doi.org/10.3847/1538-4365/ac7b8f}{\emph{The Astrophysical Journal Supplement Series} {\bfseries 262} (2022) 36}.

\bibitem{Zhang_2016}
K.~Zhang, R.~Yan, K.~Bundy, M.~Bershady, L.M.~Haffner, R.~Walterbos et~al., \emph{Sdss-iv manga: the impact of diffuse ionized gas on emission-line ratios, interpretation of diagnostic diagrams and gas metallicity measurements}, \href{https://doi.org/10.1093/mnras/stw3308}{\emph{Monthly Notices of the Royal Astronomical Society} {\bfseries 466} (2016) 3217}.

\bibitem{Sami_2018}
H.~Poetrodjojo, B.~Groves, L.J.~Kewley, A.M.~Medling, S.M.~Sweet, J.~van de Sande et~al., \emph{The sami galaxy survey: Spatially resolved metallicity and ionization mapping}, \href{https://doi.org/10.1093/mnras/sty1782}{\emph{Monthly Notices of the Royal Astronomical Society} {\bfseries 479} (2018) 5235}.

\bibitem{Biswas_2024}
P.~Biswas and Y.~Wadadekar, \emph{Structure and kinematics of star-forming elliptical galaxies in sdss-manga}, \href{https://doi.org/10.3847/1538-4357/ad4ee3}{\emph{The Astrophysical Journal} {\bfseries 970} (2024) 83}.

\bibitem{Salim_2014}
S.~Salim, \emph{Green valley galaxies}, \href{https://doi.org/10.2298/saj1489001s}{\emph{Serbian Astronomical Journal} (2014) 1–14}.

\bibitem{Xu_SFG_S0_2022}
K.~Xu, Q.~Gu, S.~Lu, X.~Ge, M.~Xiao and E.~Contini, \emph{{Star-forming S0 galaxies in the SDSS-IV MaNGA survey}}, \href{https://doi.org/10.1093/mnras/stab3013}{\emph{Monthly Notices of the Royal Astronomical Society} {\bfseries 509} (2021) 1237}.

\bibitem{dopita2013astrophysics}
M.~Dopita and R.~Sutherland, \emph{Astrophysics of the Diffuse Universe}, Astronomy and Astrophysics Library, Springer Berlin Heidelberg (2013).

\bibitem{Kewley_2019}
L.J.~Kewley, D.C.~Nicholls, R.~Sutherland, J.R.~Rigby, A.~Acharya, M.A.~Dopita et~al., \emph{Theoretical ism pressure and electron density diagnostics for local and high-redshift galaxies}, \href{https://doi.org/10.3847/1538-4357/ab16ed}{\emph{The Astrophysical Journal} {\bfseries 880} (2019) 16}.

\bibitem{fischer_tachiev_mchf}
C.F.~Fischer and G.~Tachiev, \emph{Mchf/mcdhf collection, version 2},  2014.

\bibitem{Tayal_2010}
S.S.~Tayal and O.~Zatsarinny, \emph{Breit–pauli transition probabilities and electron excitation collision strengths for singly ionized sulfur}, \href{https://doi.org/10.1088/0067-0049/188/1/32}{\emph{The Astrophysical Journal Supplement Series} {\bfseries 188} (2010) 32}.

\bibitem{Werk_2014}
J.K.~Werk, J.X.~Prochaska, J.~Tumlinson, M.S.~Peeples, T.M.~Tripp, A.J.~Fox et~al., \emph{The cos-halos survey: Physical conditions and baryonic mass in the low-redshift circumgalactic medium}, \href{https://doi.org/10.1088/0004-637X/792/1/8}{\emph{The Astrophysical Journal} {\bfseries 792} (2014) 8}.

\bibitem{Tumlinson_2017}
J.~Tumlinson, M.S.~Peeples and J.K.~Werk, \emph{The circumgalactic medium}, \href{https://doi.org/https://doi.org/10.1146/annurev-astro-091916-055240}{\emph{Annual Review of Astronomy and Astrophysics} {\bfseries 55} (2017) 389}.

\bibitem{Isobe_2023}
Y.~Isobe, M.~Ouchi, K.~Nakajima, Y.~Harikane, Y.~Ono, Y.~Xu et~al., \emph{Redshift evolution of electron density in the interstellar medium at z $\sim$ 0–9 uncovered with jwst/nirspec spectra and line-spread function determinations}, \href{https://doi.org/10.3847/1538-4357/acf376}{\emph{The Astrophysical Journal} {\bfseries 956} (2023) 139}.

\bibitem{pyneb_2015}
{Luridiana, V.}, {Morisset, C.} and {Shaw, R. A.}, \emph{Pyneb: a new tool for analyzing emission lines - i. code description and validation of results}, \href{https://doi.org/10.1051/0004-6361/201323152}{\emph{\aap} {\bfseries 573} (2015) A42}.

\bibitem{Morisset_2020}
C.~Morisset, V.~Luridiana, J.~García-Rojas, V.~Gómez-Llanos, M.~Bautista and C.~Mendoza, \emph{Atomic data assessment with pyneb}, \href{https://doi.org/10.3390/atoms8040066}{\emph{Atoms} {\bfseries 8} (2020) }.

\bibitem{PHILLIPS_2008}
J.~Phillips, \emph{Constraints upon density gradients in evolved hii regions}, \href{https://doi.org/https://doi.org/10.1016/j.newast.2007.06.013}{\emph{New Astronomy} {\bfseries 13} (2008) 60}.

\bibitem{Kronberg_2008}
P.P.~Kronberg, M.L.~Bernet, F.~Miniati, S.J.~Lilly, M.B.~Short and D.M.~Higdon, \emph{A global probe of cosmic magnetic fields to high redshifts}, \href{https://doi.org/10.1086/527281}{\emph{The Astrophysical Journal} {\bfseries 676} (2008) 70}.

\bibitem{Jelic_2015}
{Jelić, V.}, {de Bruyn, A. G.}, {Pandey, V. N.}, {Mevius, M.}, {Haverkorn, M.}, {Brentjens, M. A.} et~al., \emph{Linear polarization structures in lofar observations of the interstellar medium in the 3c196 field}, \href{https://doi.org/10.1051/0004-6361/201526638}{\emph{\aap} {\bfseries 583} (2015) A137}.

\bibitem{Filling_fraction}
E.M.~{Berkhuijsen}, D.~{Mitra} and P.~{Mueller}, \emph{{Filling factors and scale heights of the diffuse ionized gas in the Milky Way}}, \href{https://doi.org/10.1002/asna.200510488}{\emph{Astronomische Nachrichten} {\bfseries 327} (2006) 82} [\href{https://arxiv.org/abs/astro-ph/0511172}{{\ttfamily astro-ph/0511172}}].

\bibitem{astropy_2013}
{Astropy Collaboration}, T.P.~{Robitaille}, E.J.~{Tollerud}, P.~{Greenfield}, M.~{Droettboom}, E.~{Bray} et~al., \emph{{Astropy: A community Python package for astronomy}}, \href{https://doi.org/10.1051/0004-6361/201322068}{\emph{\aap} {\bfseries 558} (2013) A33} [\href{https://arxiv.org/abs/1307.6212}{{\ttfamily 1307.6212}}].

\bibitem{astropy_2018}
{Astropy Collaboration}, A.M.~{Price-Whelan}, B.M.~{Sip{\H{o}}cz}, H.M.~{G{\"u}nther}, P.L.~{Lim}, S.M.~{Crawford} et~al., \emph{{The Astropy Project: Building an Open-science Project and Status of the v2.0 Core Package}}, \href{https://doi.org/10.3847/1538-3881/aabc4f}{\emph{\aj} {\bfseries 156} (2018) 123} [\href{https://arxiv.org/abs/1801.02634}{{\ttfamily 1801.02634}}].

\bibitem{astropy_2022}
{Astropy Collaboration}, A.M.~{Price-Whelan}, P.L.~{Lim}, N.~{Earl}, N.~{Starkman}, L.~{Bradley} et~al., \emph{{The Astropy Project: Sustaining and Growing a Community-oriented Open-source Project and the Latest Major Release (v5.0) of the Core Package}}, \href{https://doi.org/10.3847/1538-4357/ac7c74}{\emph{apj} {\bfseries 935} (2022) 167} [\href{https://arxiv.org/abs/2206.14220}{{\ttfamily 2206.14220}}].

\end{thebibliography}\endgroup

\end{document}